\pdfoutput=1

\documentclass[11pt,twoside,a4paper,cmspaper,final,collab]{cms-tdr}
\def\svnVersion{58049}\def\cmsCernNo{2011-055}\def\cmsCernDate{\today}\def\cmsMessage{Submitted to the Journal of High Energy Physics}

\begin{document}\cmsNoteHeader{TOP-11-002}

\hyphenation{had-ron-i-za-tion}
\hyphenation{cal-or-i-me-ter}
\hyphenation{de-vices}
\RCS$Revision: 58012 $
\RCS$HeadURL: svn+ssh://alverson@svn.cern.ch/reps/tdr2/papers/TOP-11-002/trunk/TOP-11-002.tex $
\RCS$Id: TOP-11-002.tex 58012 2011-05-27 16:24:21Z alverson $

\providecommand{\POWHEG} {\textsc{Powheg}\xspace}

\def\mrm{\mathrm}
\def\ra{\rightarrow}

\newcommand{\roots}{\ensuremath{\sqrt{s}}}
\newcommand{\lhcE}[1]{\ensuremath{\roots ={#1}~\TeV}}

\newcommand{\FIXMEC}[1]{{\textcolor{red}{#1}}}

\renewcommand{\sign}{\ensuremath{\mathrm{sign}}}
\newcommand{\rphi}{\text{$r$-$\phi$}}
\newcommand{\etaphi}{\text{$\eta$-$\phi$}}
\newcommand{\rz}{\text{$r$-$z$}}
\newcommand{\met} {\ensuremath{E\!\!\!\!/_{\mathrm{T}}}}
\newcommand{\jpt}{\ensuremath{\mathrm{JPT}}}
\newcommand{\tcmet}{\ensuremath{\mrm{tcMET}}}
\newcommand{\calomet}{\ensuremath{\mrm{\met^{\mrm{calo,Type1}}}}}
\newcommand{\pfmet}{\ensuremath{\mrm{\met^{\mrm{PF}}}}}
\newcommand{\jet}{\ensuremath{\mrm{jet}}}
\newcommand{\jets}{\ensuremath{\mrm{jets}}}
\newcommand{\njet}{\ensuremath{N_\jet}}
\newcommand{\isotk}{\ensuremath{I_\mrm{trk}}}
\newcommand{\isocal}{\ensuremath{I_\mrm{cal}}}
\newcommand{\isocomb}{\ensuremath{I_\mrm{rel}}}
\newcommand{\mll}{\ensuremath{M_{\ell\ell}}}
\newcommand{\etsc}{\ensuremath{E_{T}^{\mrm{sc}}}}
\newcommand{\dxybs}{\ensuremath{d_0^{\mrm{BS}}}}

\newcommand{\PZ}{\ensuremath{\cmsSymbolFace{Z}}}
\newcommand{\jp}{\ensuremath{\PJgy}}
\newcommand{\eepm}{\ensuremath{\Pep\Pem}}
\newcommand{\mmpm}{\ensuremath{\Pgmp \Pgmm}}
\newcommand{\ttpm}{\ensuremath{\Pgt^+ \Pgt^-}}
\newcommand{\empm}{\ensuremath{\Pe^\pm \Pgm^\mp}}
\newcommand{\dy}{\ensuremath{\PZ/\Pgg^\star}}
\newcommand{\dyee}{\ensuremath{\dy\to\eepm}}
\newcommand{\dymm}{\ensuremath{\dy\to\mmpm}}
\newcommand{\dytt}{\ensuremath{\dy\to\ttpm}}
\newcommand{\Zee}{\ensuremath{\PZ\to\eepm}}
\newcommand{\Zmm}{\ensuremath{\PZ\to\mmpm}}
\newcommand{\wen}{\ensuremath{\PW\to \Pe\Pgne}}
\newcommand{\wmn}{\ensuremath{\PW\to\Pgm\Pgngm}}
\newcommand{\wtn}{\ensuremath{\PW\to\Pgt\Pgngt}}
\renewcommand{\ttbar}{\ensuremath{\cmsSymbolFace{t}\cmsSymbolFace{\overline{t}}}}
\newcommand{\tW}{\ensuremath{\cmsSymbolFace{t}\PW}}
\newcommand{\VV}{\ensuremath{\cmsSymbolFace{VV}}}
\newcommand{\WoZ}{\ensuremath{\PW/\PZ}}

\newcommand{\EEE}[1]{\ensuremath{\times 10^{#1}}}
\newcommand{\pp}{\ensuremath{\Pp\Pp}}
\newcommand{\ppbar}{\ensuremath{\Pp\Pap}}

\newcommand{\totLumi}{\ensuremath{36~\pbinv}}
\newcommand{\totLumiWerr}{\ensuremath{35.9\pm 1.4~\pbinv}}

\newcommand{\nNoNo}{\ensuremath{N_{\overline{n}\overline{n}}}}
\newcommand{\nNoNu}{\ensuremath{N_{{n}\overline{n}}}}
\newcommand{\nNuNu}{\ensuremath{N_{{n}{n}}}}

\newcommand{\sye}[1]{\ensuremath{~\pm #1}}
\newcommand{\ase}[2]{\ensuremath{_{~- #1}^{~+ #2}}}
\newcommand{\asi}[2]{\ensuremath{_{~- #1}^{~+ #2}}}

\newcommand{\Routin}{\ensuremath{R_\mrm{out/in}}}
\newcommand{\DRoutin}{\ensuremath{\delta(\Routin)/\Routin}}
\newcommand{\RTL}{\ensuremath{R_\mrm{TL}}}

\newcommand{\ns}{\ensuremath{\mrm{ns}}}

\cmsNoteHeader{TOP-11-002}
\title{Measurement of the \ttbar\ production cross section and the top quark mass in the
dilepton channel in pp collisions at $\sqrt{s} =7$~TeV}

\address[slava]{UC Santa Barbara}
\address[fnal]{Fermilab}
\address[cern]{CERN}
\author[cern]{The CMS Collaboration}

\date{\today}

\abstract{
The \ttbar\ production cross section and top quark mass
are measured in proton-proton collisions at $\sqrt{s} = 7\rm~TeV$
in a data sample corresponding to an integrated luminosity of $36{\rm~pb}^{-1}$ collected by the CMS experiment.
The measurements are performed in events with two leptons (electrons or muons) in the final state.
Results of the cross section measurement in events with and without b-quark identification are obtained and combined.
The measured value is $\sigma_{{\rm t\bar t}}=168 \pm 18 \,({\rm stat.})\pm 14 \,({\rm syst.}) \pm 7\,({\rm lumi.})\rm~pb$,
consistent with predictions from the standard model.
The top quark mass $m_{\rm top}$ is reconstructed with two different methods,
a full kinematic analysis and a matrix weighting technique.
The combination yields a measurement of $m_{\rm top}=175.5 \pm 4.6 \,({\rm stat.}) \pm 4.6\,({\rm syst.})\GeVcc$.
}

\hypersetup{%
pdfauthor={CMS Collaboration},%
pdftitle={Measurement of the t t-bar production cross section and the top quark mass in the dilepton channel in pp collisions at sqrt(s) =7 TeV},%
pdfsubject={CMS},%
pdfkeywords={CMS, physics, software, computing}}

\maketitle 
\newcommand{\mymet}{\makebox[2.4ex]{\ensuremath{\not\!\! E_{\mathrm{T}}}}}
\newcommand{\myttbar}{\ensuremath{{t\overline{t}}}\xspace}
\newcommand{\Bot}{\ensuremath{\cmsSymbolFace{b}}\xspace}

\section{Introduction}
\label{sec:introduction}
For many years after its discovery~\cite{top-discovery2,top-discovery3}, the properties of the top quark have been 
the subject of numerous detailed
studies~\cite{Incandela:2009pf}, which until recently have only been possible at the Tevatron proton-antiproton (\ppbar) collider.  
With the advent of the Large Hadron Collider (LHC)~\cite{lhc}, top quark processes
can now be studied extensively in multi-TeV proton-proton (pp) collisions~\cite{top10001,atlasTop3pb}.  
In both \ppbar\ and \pp\ collisions, top quarks are produced primarily in top-antitop (\ttbar) quark pairs 
via the strong interaction.  
At the LHC, the \ttbar\ production mechanism is dominated by the gluon fusion process, whereas at the Tevatron, top quark
pairs are predominantly produced through quark-antiquark annihilation.
Measurements of top quark production at the LHC are therefore important new tests of our understanding of 
the \ttbar\ production mechanism.  
The top quark mass is an important parameter of the standard model (SM) and it affects predictions 
of SM observables via radiative corrections.
A precise measurement of the top quark mass is crucial since it constitutes one of the most 
important inputs to the global electroweak fits~\cite{globalEWKfits} 
that provide constraints on the model itself, including indirect limits on the mass of the Higgs boson.
The mass of the top quark has been measured very precisely by the Tevatron experiments, 
and the current world average is $173.3\pm 0.6 \,({\rm stat.}) \pm 0.9 \,({\rm syst.})~\text{GeV}\!/\!c^2$~\cite{topmass_tev}. 
Of all quark masses, the mass of the top quark is known with the smallest fractional uncertainty. 

Within the SM, the top quark decays via the weak process ${\rm t\rightarrow \PW b}$ almost exclusively.  
Experimentally, top quark pair events are categorised according to the decay of the two \PW\ bosons: 
the all-hadronic channel, in which both \PW\ bosons decay into quarks; the lepton+jets channel, in which one \PW\
boson decays leptonically and the other into quarks; and the dilepton channel, in which both \PW\ bosons decay into leptons.  
The measurement described herein is performed using dilepton \ttbar\ modes (\eepm, \mmpm, and \empm).
These modes compose ($6.45 \pm 0.11$)\%~\cite{PDG} of
the total branching fraction for \ttbar\ when including
contributions from tau leptons that subsequently decay to electrons and muons, as is done here.
The final state studied in this analysis contains two oppositely charged leptons (electrons or muons), two neutrinos from
the \PW-boson decays, and two jets of particles resulting from the hadronization of the \Bot\ quarks.  

In this paper, a measurement of the \ttbar\ production cross section in the dilepton final state and the first 
measurement of the top quark mass in pp collisions at $\sqrt{s}=7$~TeV are described.  
The cross section analysis improves upon our previous measurement~\cite{top10001} with refined event 
selection and analysis methods, and with about twelve times more data.  
Similar measurements have been performed recently at the Tevatron~\cite{CDFdil,D0dil}
and at the LHC~\cite{atlasTop3pb}.
In addition to a measurement of the cross section, a measurement of the ratio of cross sections for 
 \ttbar\ and \dy\ production is provided.
The top quark mass is measured with two methods, a full kinematic analysis and a matrix weighting technique, 
which have been improved over those used at the Tevatron~\cite{kin,D01998}.
The results are based on a data sample corresponding to an integrated luminosity of \totLumiWerr\ 
recorded by the Compact Muon Solenoid (CMS) experiment~\cite{JINST}.

The structure of this paper is as follows:
a brief description of relevant detector components is provided in Section~\ref{sec:detector},
followed by details of the simulated samples given in Section~\ref{sec:simulation}, and
a description of data samples and event selection in Section~\ref{sec:event_selection}.
The measurement of the cross section is presented in Section~\ref{sec:xs} and the measurement of the top quark mass in Section~\ref{sec:mass}.

\section{The CMS detector}
\label{sec:detector}
The central feature of the CMS apparatus is a superconducting solenoid,
13~m in length and 6~m in diameter, which provides an axial magnetic
field of 3.8~T.  The bore of the solenoid is outfitted with various
particle detection systems.  Charged particle trajectories are
measured by the silicon pixel and strip tracker, covering $0 < \phi <
2\pi$ in azimuth and $|\eta |<$~2.5, where the pseudorapidity $\eta$ is
defined as $\eta =-\ln[\tan{\theta/2}]$, with $\theta$ being the
polar angle of the trajectory of the particle with respect to the
counterclockwise beam direction.  
A crystal electromagnetic calorimeter
(ECAL) and a brass/scintillator hadronic calorimeter surround
the tracking volume;
in this analysis the calorimetry provides high-resolution energy and
direction measurements of electrons and hadronic jets.  
Muons are measured in
gas-ionisation detectors embedded in the steel return yoke outside the solenoid.
The detector is nearly hermetic, allowing for energy balance
measurements in the plane transverse to the beam directions. 
A two-level trigger system selects the most interesting \pp\ collision
events for use in physics analysis.  
A more detailed description of
the CMS detector can be found elsewhere~\cite{JINST}.

\section{Signal cross section and event simulation}
\label{sec:simulation}
The SM expectation for the \ttbar\ production cross section at $\sqrt{s}=7$~TeV, calculated at the next-to-leading order (NLO)
using {\sc mcfm}~\cite{mcfm,mcfm:tt} for a top quark mass of 172.5\GeVcc, is $158^{+23}_{-24} \rm\ pb$.
Approximate next-to-next-to-leading-order (NNLO) calculations for the \ttbar\ cross section are also
available~\cite{Kidonakis:2010dk,Kidonakis:2008mu,Cacciari:2008zb,Moch:2008qy,Langenfeld:2009tc,Langenfeld:2009wd,Ahrens:2010zv} 
with a value of $163\ase{10}{11}$~pb, calculated for  a top quark mass of 173\GeVcc in Ref.~\cite{Kidonakis:2010dk}.
A significant part of this uncertainty is due to uncertainties on the parton distribution functions (PDFs).
These expected values can be compared to previous measurements of
${\rm 194 \pm 72 {\rm \,(stat.)} \pm 24 {\rm \,(syst.)} \pm 21 {\rm \,(lumi.)}}$~pb in events with two leptons~\cite{top10001} and
$145\pm 31 {\rm \,(stat.)} \ase{27}{42} \,{\rm (syst.)}$~pb in a combined measurement using events 
with one and two leptons~\cite{atlasTop3pb}.
The sensitivity to the PDFs  is increased in the ratio of the \ttbar\ 
and \dy\ production cross sections, which have partially anti-correlated uncertainties on theory predictions~\cite{cteqCorrelations}.

The selection efficiency of signal events is evaluated in a simulated \ttbar\ event sample modelled using 
the \MADGRAPH event generator (v.~4.4.12)~\cite{madgraph}
with matrix elements corresponding to up to three additional partons.
The generated events are subsequently processed with \PYTHIA (v.~6.422)~\cite{pythia} 
to provide the showering of the partons, and to perform the matching
of the soft radiation with the contributions from the matrix element.
Tau decays are handled with \TAUOLA (v.~27.121.5)~\cite{tauola}.
The CMS detector response is simulated using \GEANTfour (v.~9.3 Rev01)~\cite{geant}.
Events in this simulated signal sample are normalised to the NLO \ttbar\ production cross section.
In addition, for the mass measurement, different samples are generated with top quark 
masses between 151 
and $199\GeVcc$ in steps of $3\GeVcc$.

Simulated signal samples with \MADGRAPH are produced using different settings in 
order to estimate systematic 
effects on modelling of the dilepton events.
Samples are produced using different i) QCD radiation in the parton showering, ii) dynamical transferred 
four-momentum $Q^2$ event scale (varied by a factor of two, up and down), 
iii) thresholds for matching between matrix elements and  parton showers, and iv) values of the top quark mass.
Contributions from the effects of modelling the final-state particle decays are assessed
by comparing expectations derived using \PYTHIA\ alone with samples in which
the particle decays are handled by \EVTGEN~\cite{evtgenNIM} or \TAUOLA.
A sample generated with \ALPGEN~\cite{alpgen} and subsequently processed with \PYTHIA\ is used to assess differences
in the matrix element generators.
Two samples generated with \POWHEG~\cite{powheg} and subsequently processed with \PYTHIA and
 \HERWIG~\cite{herwig} are used to assess other variations in the parton showering description,
as well as to compare with an NLO event generation.
Results from these simulated signal samples are summarised in Section~\ref{sec:systematics}.

Background samples are simulated with \MADGRAPH and \PYTHIA. 
The \PW+jet contribution is checked with both generators.
The corresponding samples include only the leptonic decays of the \PW\ boson, 
and are normalised to the inclusive NNLO cross section of  $31.3\pm 1.6$~nb,
calculated using fully exclusive \PW\ and \PZ\ production ({\sc fewz}) program~\cite{fewz}.
Drell--Yan production of charged leptons in the final state
is generated with \MADGRAPH for dilepton invariant masses above 50\GeVcc, 
and  is normalised to a cross section of $3.04\pm0.13$~nb,
computed with {\sc fewz}.
The Drell--Yan events with masses between 10 and 50\GeVcc are generated with \PYTHIA.
While this sample cross section equals $12.4$~nb, these events represent only a small fraction of the total Drell--Yan contribution
 after  the analysis lepton selections.
Single top quark  production ($\pp\to\tW$) with a corresponding cross section of $10.6\pm 0.8$~pb (calculated at NLO with {\sc mcfm}) 
is simulated with \MADGRAPH.
Finally, the diboson production of \PW\PW, \PW\PZ, and \PZ\PZ, with corresponding inclusive cross sections 
of $43.0\pm1.5$~pb, $18.8\pm 0.7$~pb, and $7.4\pm 0.2$~pb
(all calculated at the NLO with {\sc mcfm}), is simulated with \PYTHIA.

Among all  the simulated backgrounds, only the \dytt, single top, and diboson (referred to as \VV, where ${\rm V}=\PW$ or \PZ)
contributions are used directly to estimate the absolute number of background events from these contributions.
All other backgrounds are estimated from control data samples.

\section{Event selection}
\label{sec:event_selection}
Proton-proton collision events used for this analysis are selected by triggers
and are then reconstructed to provide information
on electrons, muons, jets of (hadronic) particles with an optional identification
of \Bot-quark jets, and the presence of transverse momentum imbalance.
This information is used to select the final sample of events, as described below.

The events are required to have at least one good
reconstructed proton-proton interaction vertex~\cite{trkpas}
found within 24~\cm\ from the centre of the detector along the nominal beam line
and within 2~\cm\ in a direction transverse to this beam line.
Events with significant instrumental noise in the hadron calorimeters are removed.
These selection criteria have an efficiency larger than $99.5\%$ relative to events with two leptons.

\subsection{Event trigger selection}
\label{sec:triggerSel}
Events selected for this analysis are collected using lepton triggers
in which the presence of either a muon, or one or two high transverse momentum ($\pt$) electrons are required.
The muon trigger thresholds are applied to the transverse momentum $\pt$,
while for electrons the threshold is applied to the electron transverse energy $E_{\mrm T}$
(energy deposited in the ECAL projected on the  plane transverse to the nominal beam line).
For this measurement the triggers used were changed during the data taking period to adapt to the rapid 
rise in instantaneous luminosity delivered by the LHC.
Most of the data were collected with a single muon trigger threshold of 15\GeVc,
a single electron trigger threshold of 22\GeV, and a dielectron trigger threshold of 17\GeV.

The events passing all analysis selections are required to have at least two leptons with 
momentum values and quality requirements at least as restrictive as the trigger criteria.
The efficiency for triggering on  a single lepton passing all other analysis selections
 is measured in data using electrons and muons from \PZ-boson decays, and compared 
with results from the simulation.
The efficiency is measured with the tag-and-probe method~\cite{inclusWZ3pb} using two leptons with an invariant 
mass between 76\GeVcc and 106\GeVcc, 
and is found to be above 90\% (95\%) for muons (electrons).
Since the events used in this analysis are required to have only one of the two leptons satisfying
the trigger criteria, the trigger requirements are very efficient.
The efficiencies are above 97\% in the  \mmpm\ decay mode and above 99\% in the other two modes.
Based on the measured efficiencies for the trigger to select dilepton events,
the simulated trigger efficiency is corrected by simulation-to-data scale factors of
$0.983\pm 0.007$, $1.000 \pm 0.001$, and $0.994 \pm 0.003$ for the \mmpm, \eepm, and \empm\ final states, respectively.
The uncertainties have  statistical and systematic contributions, including variations due to differences 
in lepton kinematics between the \ttbar\ signal and \PZ-boson events.

\subsection{Lepton selection}
\label{sec:leptonSel}
Energetic muons and electrons reconstructed in the event are used for the analysis.
At least two leptons in the event are required to pass identification and isolation requirements.
The selection criteria are very close to those in~\cite{top10001}.

Muon candidates are reconstructed~\cite{MUOPAS} using two algorithms
that require consistent hits in the tracker and muon systems: one matches
the extrapolated trajectories from
the silicon tracker to hits in the muon system (tracker-based muons);
the second performs a global fit of
consistent hits in the tracker and the muon system (globally fitted muons).

Electron candidates are reconstructed~\cite{EGMPAS} starting from a
cluster of energy deposits in the crystals of the ECAL, which is then
matched to hits in the silicon tracker and used to initiate a track reconstruction algorithm.  
The electron reconstruction algorithm takes into account the possibility of significant energy loss of the
electron through bremsstrahlung as it traverses the material of the tracker.  
Anomalous signals corresponding to particles occasionally interacting in the 
ECAL transducers are rejected during the reconstruction step.

The leptons are required to have $\pt>20\GeVc$ and $|\eta|<2.4$~(2.5) for muons (electrons).
The lepton candidate tracks are required to originate from near the interaction region (i.e., the beam spot):
the distance of closest approach in the transverse plane to the beam line must be less than $200~\mum$  ($400~\mum$),
and the distance between the point of closest approach to the beam line and a primary
vertex must be less than 1~\cm\ along the beam direction.

Additional quality requirements are applied to the muons. 
The track associated with the muon candidate is required to have a minimum 
number of hits in the silicon tracker, and to have a high-quality global fit 
including a minimum number of hits in the muon detector.

Several quality criteria are applied to the electron candidates.
Requirements on the values of electron identification variables based
on shower shape and track-cluster matching are applied to the
reconstructed candidates; the criteria are optimised in simulation for
inclusive \wen\ events
and are designed to maximise the rejection of electron candidates from
QCD multijet production, while maintaining 90\% efficiency for
electrons from the decay of \WoZ\ bosons.
Electron candidates within $\Delta R=\sqrt{(\Delta\phi)^2+(\Delta\eta)^2}<$~0.1 of a
tracker-based or globally fitted muon are rejected to remove 
the contribution from muon inner bremsstrahlung 
(collinear final-state radiation), where the muon track and the collinear photon are reconstructed as an electron.
Electron candidates consistent with photon conversions are rejected
based on either the reconstruction of a conversion partner in the silicon tracker,
or based on the absence of hits in the pixel tracker that are expected along the electron
trajectory originating in the collision region.

Both electron and muon candidates are required to be isolated relative to other activity in the event.
For selected muon and electron candidates, a cone of $\Delta R <$~0.3 is constructed
around the candidate's direction. In this cone, the scalar sum of the 
transverse momenta of the tracks and the calorimeter energy deposits, 
projected onto the plane transverse to the beam, is calculated.
The contribution from the candidate lepton is not included. 
The ratio of this scalar sum over the candidate's transverse momentum
defines the relative isolation discriminant, \isocomb. 
The candidate is considered to be non-isolated and is rejected if \isocomb~$>0.15$.  

The performance of the lepton candidate selection is measured using the tag-and-probe method 
in \PZ-boson events.
The electron and muon reconstruction efficiency is greater 
than 99\%~\cite{EGMPAS,wzPAS2010};
the efficiency of the quality requirements is approximately 99\% for muons and 
in the range of 85\% to 95\% for electrons;
both are reproduced well in simulation.
The average lepton isolation selection efficiency measured in real \PZ-boson events 
of 99\% (98\%) for electrons (muons) can be compared to the value of approximately 95\% from
simulated \ttbar\ signal events.
Based on an overall comparison of the muon (electron) selection efficiency in data and simulation,
the event yield selected in simulation is corrected by $0.992 \pm 0.005$  ($0.961 \pm 0.009$) per muon (electron),
where the correction also accounts for differences in the isolation and charge requirements between data and simulation.

\label{sec:dilSel}
\label{sec:btagSel}
Events are required to have at least one pair of oppositely charged leptons. 
The efficiency of this requirement depends directly on the performance of the lepton charge identification.
The muon charge misidentification is negligibly small.
The average electron charge misidentification is 0.8\%, being 0.5\% for electron tracks hitting the ECAL barrel
and up to 2\% for the ECAL endcaps.
These values are well reproduced in the simulation.

Dilepton candidate events with an invariant mass $\mll <12\GeVcc$ are removed, with
essentially no reduction in the \ttbar\ signal; this requirement suppresses dilepton pairs from heavy-flavour
resonance decays, as well as low-mass \dy\  Drell--Yan processes. 
In events with multiple pairs of leptons passing all of the requirements described so far,
only the pair of leptons with the highest transverse momenta is used for further consideration.
To veto contributions from \PZ-boson production, the
invariant mass of the dilepton system is required to be
outside the range 76 to 106\GeVcc for the \eepm\ and \mmpm\ modes.
This invariant mass requirement rejects about 90\% of \dy\ events,
at the cost of rejecting approximately 23\% of the \ttbar\ signal.

\subsection{Jet selection and \Bot-jet tagging}
Dilepton \ttbar\ events contain hadronic jets from the hadronization of the two \Bot\ quarks. 
The anti-$k_T$ clustering algorithm~\cite{antikt} with $R=0.5$ is used for jet clustering.  
Jets are reconstructed based on the calorimeter, tracker, and muon system information
using the particle flow reconstruction~\cite{JETPAS}
which provides a list of particles for each event. 
Muons, electrons, photons, and charged and neutral hadrons are reconstructed individually.
Jet energy corrections, generally smaller than 20\%, are applied to the raw jet 
momenta to establish a relative response of the calorimeter uniform as a function of the jet $\eta$, and an
absolute response uniform as a function of the jet $\pt$~\cite{JESPAS}. 
The corrections are derived using simulated events and measurements with dijet and photon+jet events.
Jet candidates are required to have $\pt > 30\GeVc$, $|\eta|<2.5$, and must not
overlap  either of the selected lepton candidates within $\Delta R< 0.4$.

Events with at least two jets 
provide the sample with the best signal-to-background ratio
for the cross section measurement,
while events with only one jet improve the acceptance and are treated separately.
Furthermore, two jets are necessary for reconstruction of the top quark candidates, 
and only such events are used for the mass measurement.
More than 95\% of \ttbar\ events have at least one jet passing the selection criteria, and approximately
three quarters of these events have at least two jets, as estimated in simulation.

The use of b tagging in the event selection can
further reject  background events without \Bot\ jets.
Furthermore, the fraction of jets correctly associated with the top quark candidates for the mass reconstruction 
can be increased significantly by using the information provided by \Bot\ tagging.
In about three quarters of the signal events with at least two jets, both \Bot-quark jets from 
the \ttbar\ decays are expected to pass the jet selection criteria.

\begin{figure}[h!]
\begin{center}
  \includegraphics[width=0.325\textwidth]{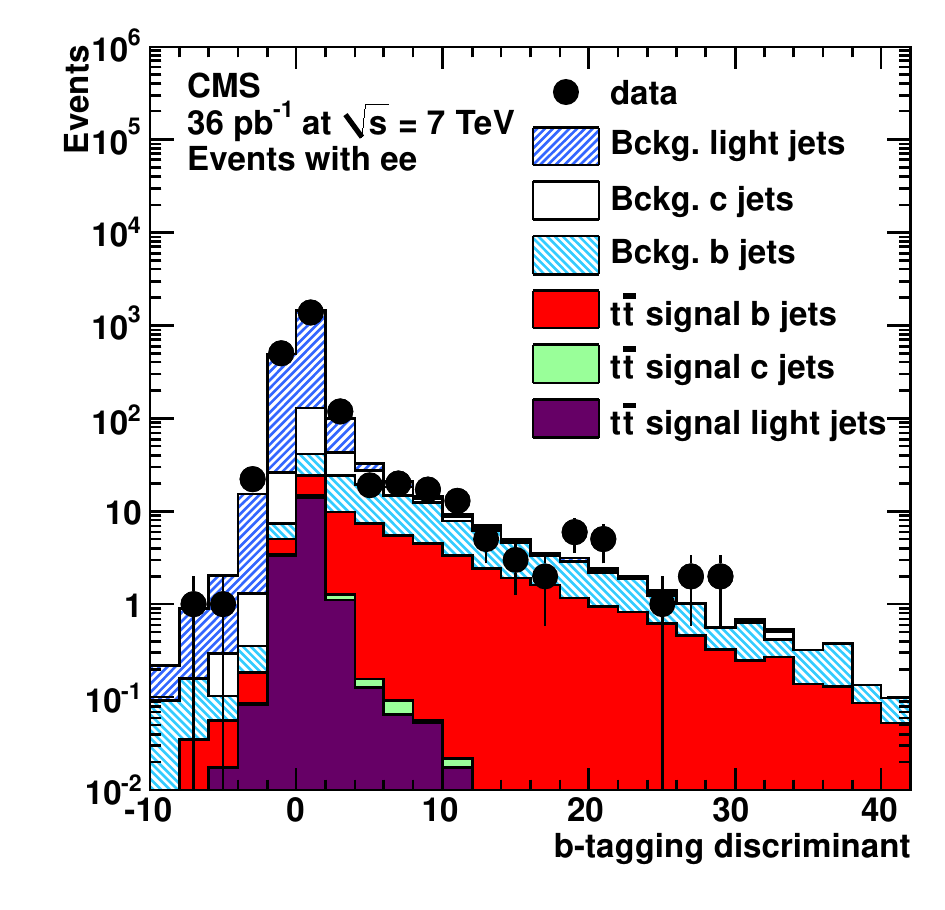}
  \includegraphics[width=0.325\textwidth]{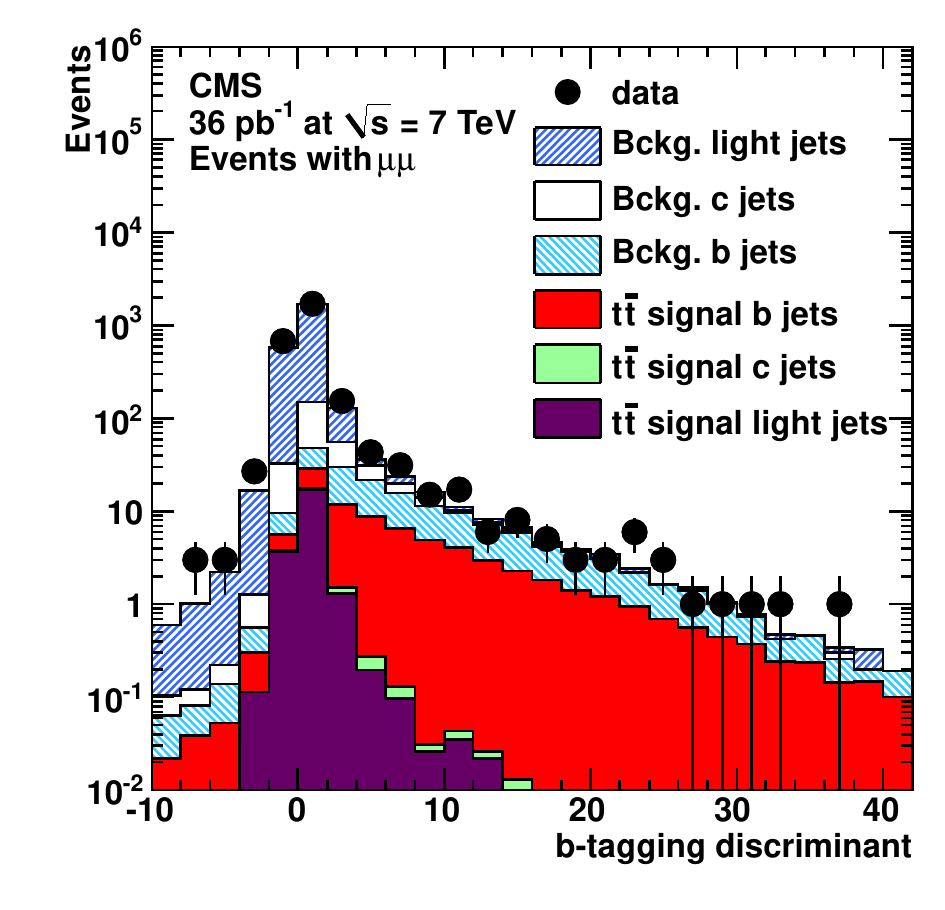}
  \includegraphics[width=0.325\textwidth]{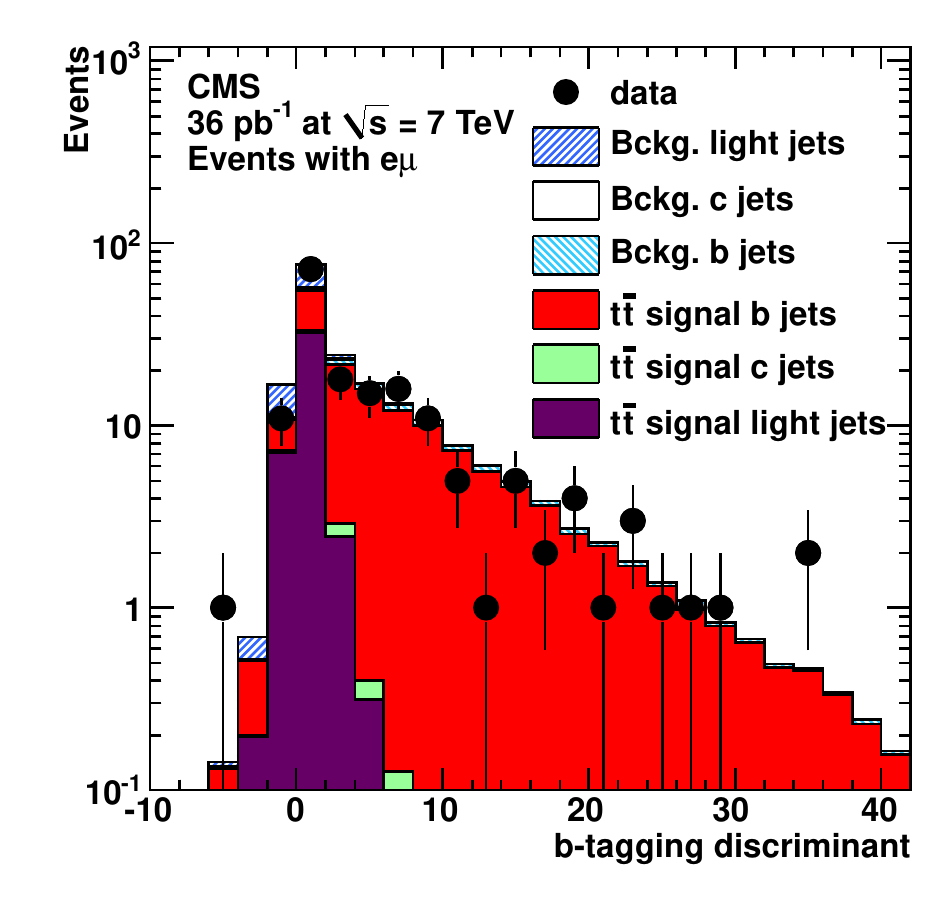}
\caption{Distribution of the \Bot-tagging discriminant in events with at least one jet
and two oppositely charged leptons in data (points),
compared to signal and background expectations from simulation (histograms)
for \eepm\ (left), \mmpm\ (centre), and \empm\ (right).
The simulated contributions are normalised to the SM predicted values without additional corrections.
All background contributions are combined and displayed separately, based on the flavour
of the simulated jet.
\label{fig:btagflavor}}
\end{center}
\end{figure}

A \Bot-quark jet identification algorithm that relies on the presence of charged
particle tracks displaced from the primary \pp\ interaction
location, as expected from the decay products of long-lived \Bot\ hadrons~\cite{BTVPAS}, is used in this analysis.
A jet is identified to be from a \Bot\ quark if it contains at least two tracks with an impact parameter significance, 
defined as the \Bot-tagging discriminant, above 1.7.
This corresponds to an efficiency of about 80\% for a \Bot-quark jet in dilepton \ttbar\ signal
events and to  a 10\% mistagging rate of light-flavour or gluon jets, as estimated in
simulation.
Good agreement is found for the distribution of this discriminant in data and simulation,
as shown in Fig.~\ref{fig:btagflavor};
a higher value corresponds to a sample with a higher fraction of genuine \Bot\ jets.
The relationship between the \Bot-tagging efficiency and the multiplicity
of the \Bot-tagged jets in the signal sample can be used to measure
the \Bot-tagging efficiency in data, as discussed in Section~\ref{sec:btagSyst}.

The \Bot-tagging procedure is used differently in the cross section and mass measurements.
For the cross section, independent measurements are made using events with and without at least one \Bot-tagged jet.
The use of \Bot\ tagging in the mass measurement is described in Section~\ref{sec:mass}.

\subsection{Missing transverse energy selection}
\label{sec:metsel}
\begin{figure}[h!]
\begin{center}
  \includegraphics[width=0.325\textwidth,viewport=0 0 590 600,clip]{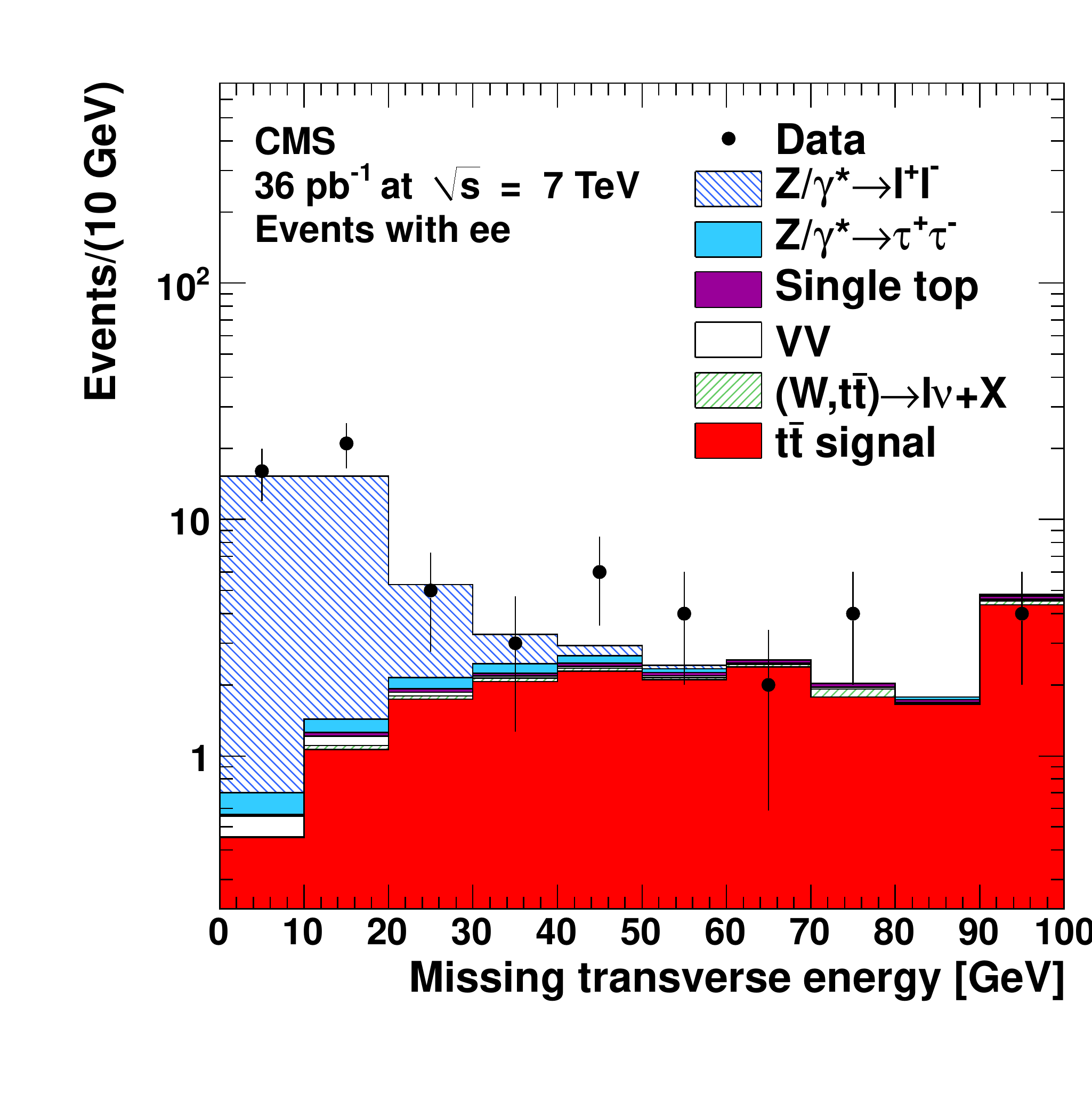}
  \includegraphics[width=0.325\textwidth,viewport=0 0 590 600,clip]{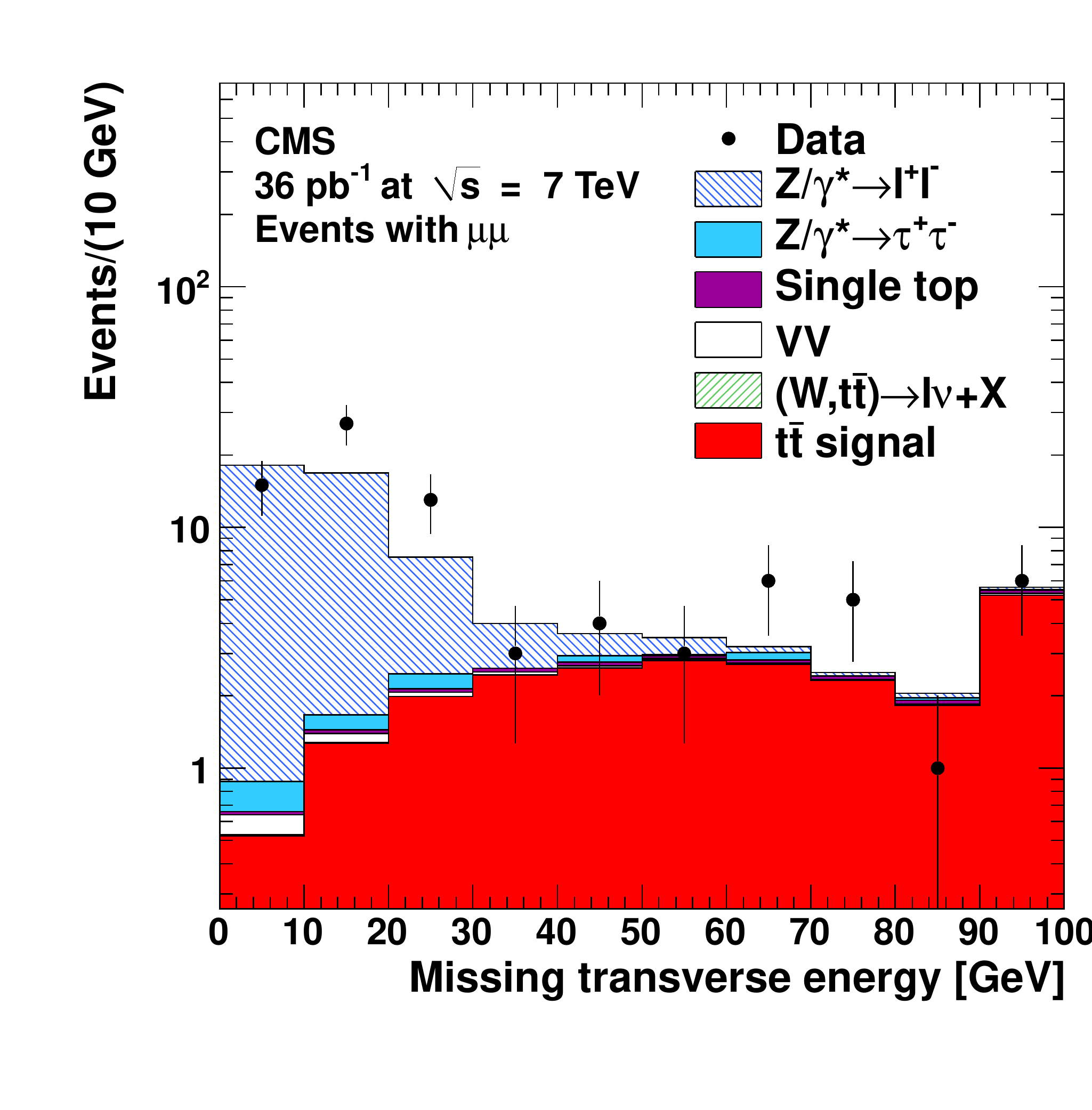}
  \includegraphics[width=0.325\textwidth,viewport=0 0 590 600,clip]{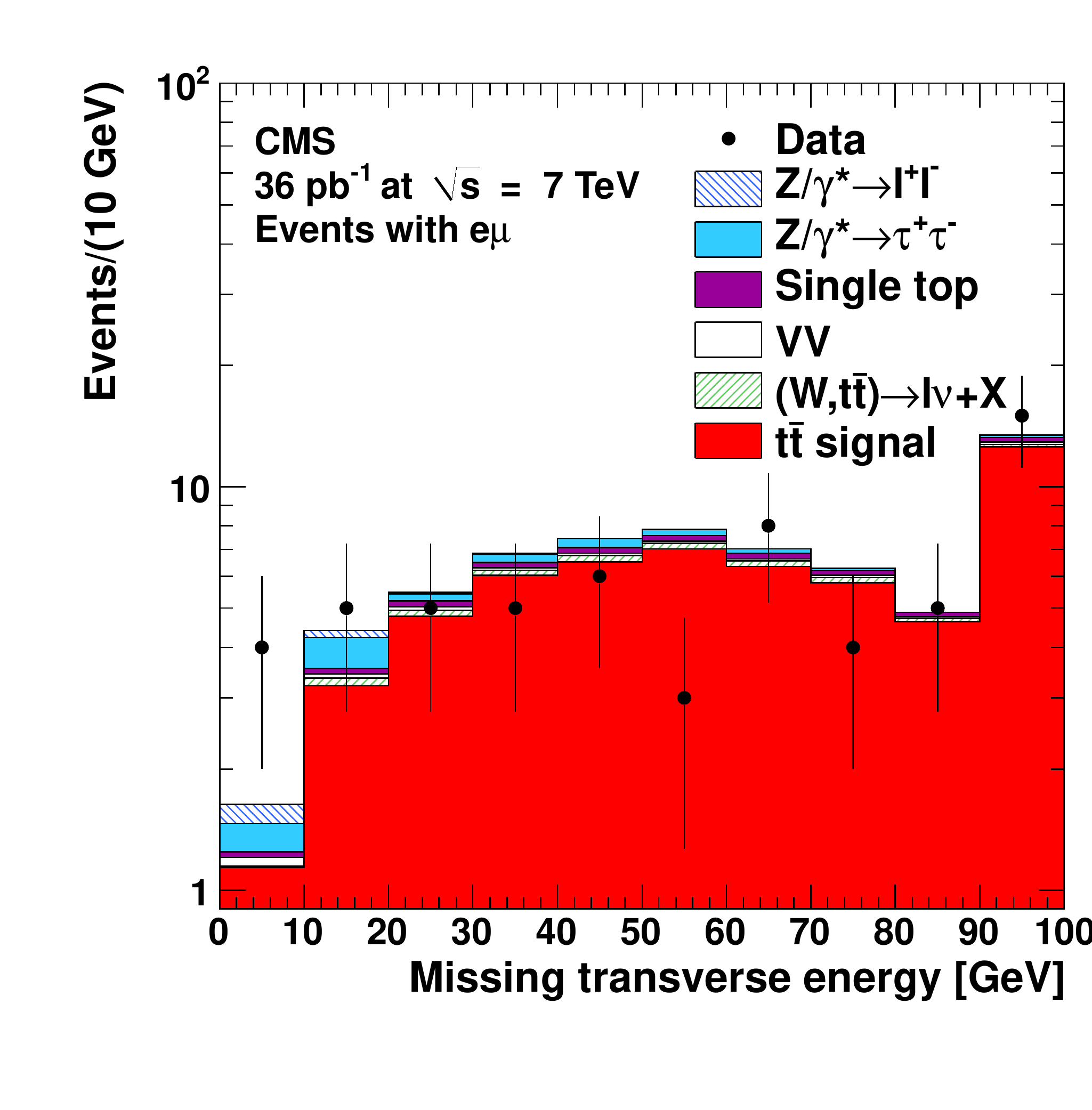}
\caption{Distribution of \met\ for events with at least two selected jets and passing the
full  dilepton selection criteria without \Bot\ tagging, except for the \met\ requirement
for \eepm\ (left), \mmpm\ (centre),  and \empm\ (right) from data (points).
The signal and background predictions from simulation are shown as the histograms.
The last bin includes the overflow contribution.\label{fig:metNM2}}
\end{center}
\end{figure}

The presence of neutrinos from the \PW-boson decays manifests itself as an imbalance in the
measured momenta of all particles' \pt, in the plane perpendicular to the beam line.
The missing transverse energy vector  $\vec{\met} = - \sum_i{c \vec{p}_{T_i}}$, and its magnitude (\met), 
are important distinguishing features of \ttbar\ events in the dilepton channel.  
The $\vec\met$ is calculated using the particle flow algorithm~\cite{METPAS2}.
The distributions of \met\ for events with at least two jets are shown in Fig.~\ref{fig:metNM2} 
(no simulation-to-data corrections are applied here).
Events selected with only one jet have a larger background contribution compared to those with at least two jets.
The missing transverse energy selection is optimised separately for these events.
The figure of merit used in the optimisation is  the expected uncertainty on the measured cross section.
It is based on a simplified model of the uncertainty on the final measurement in a given channel,
and accounts for statistical and systematic uncertainties on the signal and backgrounds.

Neither the dominant background processes, Drell--Yan \dyee\ and \mmpm, nor the
background from isolated lepton candidates produced in QCD multijet
events, contains a natural source of large $\met$.  
Hence, in the \eepm\ and \mmpm\ modes, $\met>30\GeV$ (50\GeV) is required
in events with at least two jets (only one jet)  at a loss of approximately one sixth (one third) of signal events.

For the cross section measurement, no missing transverse energy requirement is applied in the \empm\ mode, 
since the background contributions are already found to be sufficiently low.
Events with only one jet in the \empm\ final state have, however, a significant contribution from $\dytt$ background.
In order to suppress this background, these events are required to
satisfy the condition $M_T^{\Pe} + M_T^{\Pgm} > 130\GeVcc$,
which suppresses the \dytt\ by a factor of more than a hundred to a negligible level,
 at the cost of losing approximately one third of signal events.
For each lepton $\ell$ (either electron or muon), the transverse mass $M_T^{\ell}$ is defined relative 
to the transverse momentum $\pt^\ell$ 
and azimuthal direction $\phi_\ell$ of the leptons, 
and  the magnitude and the direction ($\phi_{\vec{\met}}$) of $\vec{\met}$, as 
$M_T^{\ell} = \sqrt{2 \pt^{\ell} \met [1-\cos(\phi_{\vec{\met}} - \phi_{\ell})]/c^3}$.

\section{Measurement of the cross section}
\label{sec:xs}

\subsection{Background estimates}
\label{sec:bkgd}
Two types of background estimation techniques are used in this analysis.
Backgrounds from processes expected to be small and/or simulated reasonably well
are estimated from the simulated samples described in Section~\ref{sec:simulation}.
This includes contributions from \dytt, single top, and diboson production processes.
These processes contribute events with genuine isolated leptons and genuine missing transverse energy
from the neutrinos present in the final states.
This similarity to the \ttbar\ signal events and the relatively small size of these contributions
justifies the use of simulation.
There are, however, backgrounds that are not expected to be modelled accurately.
In such cases, yields from these processes are estimated with methods using data.
One method is used to account for contributions from \dymm\ and \dyee.
Another method is used to account for events with at least one of the lepton candidates  arising from jets
misidentified as isolated lepton candidates from \PW\ or \PZ\ decays (non-\WoZ\ lepton candidates).

\subsubsection{Events from \dyee\ and \mmpm}
\label{sec:dydata}
The number of events $N_{\dy}^{\rm out}$  from Drell--Yan $\dyee$ and \mmpm\ in the  sample of events passing the \PZ-boson veto
is estimated using the method described in~\cite{top10001}.
This contribution is derived from the number of \dyee\ and \mmpm\ data events with a dilepton 
invariant mass $76< \mll<106\GeVcc$, 
scaled by the ratio of events failing and passing this selection estimated in simulation ($\Routin$).
The number of \eepm\ and \mmpm\ Drell--Yan events near
the \PZ-boson peak $N_{\dy}^{\rm in}$ is given by the number of all
events failing the \PZ-boson veto $N^{\rm in}$ after subtraction of the non-Drell--Yan
contribution.
The non-Drell--Yan contribution is estimated from \empm\ events passing
the same selection $N_{\empm}^{\rm in}$  and corrected for the differences between the
electron and muon identification efficiencies $k$.
The \dy\ contribution is thus given by
\begin{eqnarray*}
N_{\dy}^{\rm out} = \Routin N_{\dy}^{\rm in}  =  \Routin (N^{\rm in} - 0.5 k N_{\empm}^{\rm in}).
\end{eqnarray*}
The correction $k$ is estimated from $k^2=N_{\eepm}/N_{\mmpm}$ for the \dyee\ contribution 
and from $k^2=N_{\mmpm}/N_{\eepm}$ for the \dymm\ contribution, where $N_{\eepm}$ ($N_{\mmpm}$)
is the number of dielectron (dimuon) events
near the \PZ-boson mass, without a requirement on the missing transverse energy.

The systematic uncertainty on the predictions of this method is dominated by the uncertainty on \Routin.
The value of \Routin\ is estimated in simulation; 
it is found to be affected by the detector calibration effects
and to change significantly with increasingly stringent requirements on \met\ and jets in the event.
The systematic uncertainty is estimated from these variations.
The missing transverse energy requirement on selected events corresponds to an enhancement in the fraction of leptons with
mismeasured momenta, which directly contributes to an increase in \Routin.
This increase is most significant for dimuon events and 
contributes 30\% to 50\% of the total systematic uncertainty;
the increase for electrons is less significant and is less than 20\%.
The energy scale calibration effects contribute 
approximately 15\% in dielectron events but are not significant for muons.
The requirement on the presence of jets broadens the dilepton invariant mass line shape,
leading to an additional uncertainty of 15\%.
Statistical uncertainties on these estimates in simulation are 20\%.
The combined systematic uncertainty of this method, evaluated in each mode separately, 
is estimated to be 50\%.  

The estimates of the \dyee\ and \dymm\ contributions 
are given at the end of Section~\ref{sec:results_channels}.
The statistical uncertainties of these estimates are approximately equal to the systematic uncertainties.

\subsubsection{Events with leptons from non-W/Z decays}
\label{sec:fakesdata}
Background contributions with at least one non-\WoZ\ lepton candidate are expected to arise 
predominantly from multijet and \PW+jet events as well as from \ttbar\ events, 
with at most one \PW\ boson decaying leptonically.
Based on simulation, events with non-\WoZ\ lepton candidates passing the final signal selections
are expected to have similar contributions from \ttbar\ and \PW+jet events
with the fraction of \ttbar\ events increasing after the b-tagging requirement.
Simulation is not expected to predict all contributions with non-\WoZ\ lepton candidates.
Estimates on these backgrounds are derived from data.

The number of events with non-\WoZ\ leptons is estimated using a sample of
dilepton candidates that pass looser lepton identification criteria, but fail the full
selections.
The fraction of lepton candidates from non-\WoZ\ leptons passing the full selection 
relative to those passing the loosened criteria is defined as the tight-to-loose ratio
\RTL.
It is expected to be approximately independent of the sample in which
 the non-\WoZ\ lepton candidate is found,
based on observations in simulation and in data.
We measure \RTL\  using a data sample dominated by multijet events (\RTL\ calibration sample), selected
in a sample with a single loose lepton candidate,
with additional requirements vetoing events with significant transverse momentum 
consistent with \PW-boson production,
or with another lepton consistent with \PZ-boson production.

Different choices of looser selections are considered.
The isolation requirements of $\isocomb<0.4$ and $\isocomb<1.0$ are used for muons separately.
Selections with looser identification 
(no requirement on calorimeter cluster shape or cluster-to-track matching information)
 and, separately, a looser  isolation ($\isocomb<1.0$) are used for electrons.
The measured value of \RTL\ changes slightly as a
function of candidate \pt\ and $|\eta|$ for both muon and electron candidates.
The value of \RTL\ is similar for electrons and muons, and is in the range of 0.2 to 0.4 (0.02 to 0.05) 
for loose (looser) lepton selection.
Extensive tests were performed to confirm that these choices of looser lepton selection
criteria yield measurements of \RTL\  appropriate for use in the
dilepton signal sample.
These tests were done using simulated samples, as well as data events with same-sign lepton pairs,
which are dominated by non-\WoZ\ lepton candidates.
Measurements of \RTL\ using these two different definitions were subsequently combined using 
a simple mean of central values and taking the larger uncertainty as a conservative estimate.

The number of background events with one and two  non-\WoZ\ lepton candidates
is derived separately 
using a sample of dilepton events 
with both leptons failing the tight selection criteria,
and a sample with only one lepton failing.
The signal contamination in these  samples is subtracted by
taking the number of events with two leptons passing the tight selection and scaling by an
efficiency correction factor derived from a sample of \PZ\ 
 events passing the looser selection, but with the same jet multiplicity requirement.

The systematic uncertainties on the number of background events with non-\WoZ\ lepton candidates
are  primarily from the estimate of \RTL.
They  arise from differences in the momentum spectrum and flavour composition 
between the \RTL\ calibration sample and the sample where it is applied.  
The uncertainty due to momentum spectrum differences is about $60\%$ for muons and $25\%$ for electrons.
The uncertainty due to the flavour composition differences is approximately $20\%$ for both 
muons and electrons.
Other smaller contributions include those from the electroweak signal contribution 
(approximately $20\%$ for muons and negligible for electrons), 
differences in the event trigger
selections between the \RTL\ calibration sample and the signal sample to
which it is applied (generally within $20\%$ in addition to already accounted effects), 
and from the statistical limitations on the \RTL\ calibration sample.  
The systematic uncertainty on the electron (muon) \RTL\ is 50\% (75\%), which corresponds to a 50\% (75\%) systematic uncertainty
on the estimate of events with one non-\WoZ\ isolated lepton and  100\% for events with two such candidates.
The final estimate of the non-\WoZ\ contribution also includes a systematic uncertainty
on the signal contamination to the background samples, equal to about $1\%$ of the total 
signal contribution.
This is estimated from the observed variation in the contamination rate as a function of the number
of jets in  \PZ-boson events.

Results of estimates of the number of events with  non-\WoZ\ lepton candidates
are summarised at the end of Section~\ref{sec:results_channels}.
In all cases the statistical uncertainties on the estimates are comparable to or larger than the systematic
uncertainties.
There is a reasonable agreement between the number of  events expected from the simulation 
and these estimates from data.

\subsection{Systematic effects}
\label{sec:systematics}
Systematic uncertainties and corrections considered in this measurement
are from uncertainties and biases in the detector performance, from
variations in the signal acceptance due to imperfect knowledge of the
signal production, from background estimates, and from the absolute
normalisation of the integrated luminosity (4\%)~\cite{lumipas}.

\subsubsection{Selection of leptons}
\label{sec:leptonSyst}
The rates of events selected in the simulated signal sample are corrected based on comparisons
of single-lepton selection efficiencies in data and simulation using \PZ-boson events as mentioned in Section~\ref{sec:event_selection}.
The simulation-to-data scale factors with uncertainties including statistical and systematic contributions are
$SF^{\Pe\Pe} = 0.923 \pm 0.018$ in the dielectron final state, $SF^{\Pgm\Pgm} = 0.967 \pm 0.013$ in the dimuon
final state, and  $SF^{\Pe\Pgm} = 0.947 \pm 0.011$ in the electron-muon final state.
The dielectron and dimuon scale factors are not correlated with respect to each other,
while the correlation coefficient of $SF^{\Pe\Pgm}$ is approximately 0.83 and 0.56
with the dielectron and dimuon scale factors, respectively.

The electron and muon isolation selection efficiency is about 4\% lower
per lepton in simulated \ttbar\ events compared to \PZ-boson events passing the same requirements on the jet multiplicity.
A fractional uncertainty of 50\% is assigned to the overall effects responsible for this difference,
corresponding to an additional uncertainty of 2\% per lepton (4\% per event) attributed to
the lepton selection modelling.

The lepton momentum scale is known to better than 1\% for muons and  electrons in the barrel ECAL,
and to approximately 2.5\% for electrons found in the endcap part of the ECAL, based on
comparisons of the position of the \PZ-boson mass peak in data to its value in simulation.
The effect of the bias in the electron energy in the ECAL endcap is included in the simulation-to-data scale
factor shown above.
The uncertainty on the \ttbar\ selection due to the momentum scale is estimated
 to be less than 1\% and is neglected.

\subsubsection{Selection of jets and missing transverse energy}
\label{sec:jesSyst}
The uncertainty on the jet energy scale is directly related to the efficiency of jet 
and missing transverse energy selection.
The effect of the jet scale uncertainty is estimated from the change
of the number of selected simulated \ttbar\ events by simultaneously varying jet momenta up or down
within the uncertainty envelope of the jet energy scale, corresponding to one standard deviation.
This envelope corresponds to a combination of the following:
the inclusive jet scale uncertainty estimated from data~\cite{JESPAS} to be 
in the range of 2.5--5\% (dependent on jet \pt\ and $\eta$);
a contribution of 1.5\% to account for differences in the reconstruction
 and simulation software in~\cite{JESPAS} and here;
and an uncertainty of $2\%$ to $3\%$ (dependent on transverse momentum) corresponding to the difference in response between inclusive and b-quark jets in \ttbar\ events.
Variations in the jet momenta are propagated to the value of the 
missing transverse energy in this procedure.
In addition, the remaining small fraction of the missing transverse energy
that is not associated to measurements of  jets or leptons
is varied by 10\% independently of the jet scale variation to account for an uncertainty on the missing
transverse energy from the unclustered hadronic contribution.
The systematic uncertainty attributed to the hadronic energy scale 
(the combined effect of the jet and missing transverse
energy scales) is estimated separately for each selection, 
averaged over the \eepm\ and \mmpm\ final states, and
separate from the \empm\ final state as summarised in Table~\ref{tab:systSumm}.
The uncertainty on the number of events with one jet is anti-correlated with the uncertainty 
on the number of events with at least two jets.
The systematic effects due to differences in the jet energy resolution in data and simulation
are found to be negligible.

The effect arising from  the presence of  additional proton-proton collisions (pileup) is estimated separately.
Lepton selection simulation-to-data scale factors and uncertainties described in Section~\ref{sec:leptonSyst}
naturally include the contribution from pileup.
The remaining effect is on the jet and missing transverse energy selection: it
introduces a small bias by increasing the number of selected jets.
The corresponding scale factor applied to simulation due to pileup effects
is $1.013\pm 0.008$ in events with at least two jets, and $0.967\pm 0.020$ in events
with only one jet.

\label{sec:btagSyst}
The uncertainty on the number of events selected with at least two jets
and  at least one b-tagged jet is estimated from data.
Neglecting the residual contribution from misidentification of light-flavour quark, c-quark,
and gluon jets present in the \ttbar\ signal sample, 
the variation in the b-tagging efficiency corresponds to
the variation of the ratio of events with at least two b-tagged jets relative 
to the number of events with at least one b-tagged jet, $R_{2/1}$, 
and the relative variation in the number of events with at least one b-tagged jet, $\frac{\delta N_1}{N_1}$.
These values are found to be in a simple relationship
$\frac{\delta N_{1}}{N_{1}} \approx 0.5 \delta R_{2/1}$.
There are 51 events with at least one b-tagged jet observed in data for the \empm\ final state
with $3.0\pm1.4$ background events expected, as described in Section~\ref{sec:yields};
30 of these events have at least two b-tagged jets 
with $0.9\pm 0.5$ background events expected.
These numbers give a value of $R_{2/1}^{\rm data} = (60.8 \pm 7.5)\%$,
to be compared to the value of $R_{2/1}^{\rm sim} = (57.9\pm0.1)\%$ from simulation,
where the uncertainty in simulation is dominated by an estimate of misidentification of light-flavour
quark and gluon jets present.
Because of the agreement of these two measurements, we make no further corrections to
the value of the efficiency to select at least one b-tagged jet in events with at least two jets.
The systematic uncertainty on this efficiency is conservatively estimated at 5\%,
derived from the measured uncertainty of $R_{2/1}^{\rm data}$, and an additional
 uncertainty of approximately $0.3\%$ on the contribution from light-flavour quark and gluon
jets.

\subsubsection{Signal modelling effects}
\label{sec:generator_systematics}
Several effects contribute to the systematic uncertainties in the modelling of the \ttbar\ production.
Only significant effects are assigned a nonzero systematic uncertainty.
In addition to the uncertainties, a correction is applied to the simulated signal sample to account for the 
leptonic branching fractions of the \PW\ boson.
The leading-order value of 1/9 set by the event generator is corrected
to match the measured value of $0.1080\pm 0.0009$~\cite{PDG}.

Systematic uncertainties on the signal event selection efficiency are included, 
as shown in Table~\ref{tab:systSumm}. 
These are based on studies of the samples described in Section~\ref{sec:simulation}:
from tau-lepton and hadron-decay modelling; event $Q^2$ scale;
a conservative uncertainty on the top quark mass (taken as 2\GeVcc);
jet and \met\ model uncertainty from comparisons between the 
matrix element generators \ALPGEN, \MADGRAPH, and \POWHEG;
and from the uncertainty in the showering model, estimated from the difference between \HERWIG and \PYTHIA.
Uncertainties on the presence of additional hadronic jets
produced as a result of QCD radiation in the initial and final states and uncertainties on the parton
distribution functions were found to have a negligible effect. 

\subsubsection{Summary of systematic effects on the signal selection}
\label{sec:systematics_summary}
Fractional uncertainties on the signal efficiency described earlier in this section
for events passing the full signal event selection are summarised in Table~\ref{tab:systSumm},
listed in the order they appear in the text.
All uncertainties are common for \eepm\ and \mmpm\ final states,
except for the uncertainty on the lepton selection.
Scale factors, which account for all known discrepancies between data
and simulation, are  applied to the simulated signal sample.
The product of the scale factors described earlier in this section are $0.883$, $0.926$, and $0.906$ 
($0.843$, $0.884$, and $0.866$)
for  the \eepm, \mmpm, and \empm\ final states, respectively, in events with at least two (only one) jets.

\begin{table}[h]
\begin{center}
\caption{\label{tab:systSumm}Summary of the relative systematic uncertainties on the
number of signal \ttbar\ events after the full selection criteria,
shown separately for each of the dilepton types and for events with only one
and more than one jet.
All values are in percent.
Systematic uncertainties on the lepton selection are treated separately 
for \eepm\ and \mmpm\ final states.
Different sources (values in different rows) are treated as uncorrelated.
Lepton selection uncertainties are correlated only in the same dilepton final state.
All other uncertainties are 100\% (anti)correlated among any two columns 
for the same source, as reported with the (opposite) same sign.
The subtotal values are for sums in quadrature of all corresponding values in the same column.
}
\begin{tabular}{lcc|cc}\hline\hline
		& \multicolumn{2}{c|}{$\njet=1$}	& \multicolumn{2}{c}{$\njet\geq 2$} \\ 
Source		& $\eepm+\mmpm$		& $\empm$	& $\eepm+\mmpm$         & $\empm$     \\ \hline
Lepton selection & 1.9/1.3     	 & 1.1                  & 1.9/1.3             & 1.1           \\
Lepton selection model & 4.0     & 4.0                  & 4.0                 & 4.0           \\
Hadronic energy scale     & $-3.0$        & $-5.5$               & $3.8$               & 2.8           \\
Pileup           & $-2.0$        & $-2.0$               & 0.8                 & 0.8           \\
\Bot\ tagging ($\geq 1$ \Bot\ tag)&      &                      & 5.0                 & 5.0           \\ \hline
Branching ratio  & 1.7           & 1.7                  & 1.7                 & 1.7           \\ 
Decay model      & 2.0           & 2.0                  & 2.0                 & 2.0           \\
Event $Q^2$ scale& 8.2           & 10                   & $-2.3$              & $-1.7$        \\
Top quark mass   & $-2.9$        & $-1.0$               & 2.6                 & 1.5           \\
Jet and \met\ model& $-3.0$      & $-1.0$               & 3.2                 & 0.4           \\
Shower model     & 1.0           & 3.3                  & $-0.7$              & $-0.7$        \\  \hline
Subtotal without \Bot\ tagging& 11.2/11.1& 13.1                 & 8.0/7.9             & 6.2           \\
Subtotal with \Bot\ tagging&              &                      & 9.5/9.4             & 8.0           \\ \hline
Luminosity	& 4.0		 & 4.0			& 4.0 		      & 4.0	      \\ \hline\hline
\end{tabular}
\end{center}
\end{table}

\subsubsection{Systematic effects on background estimates}
\label{sec:background_systematics}
Uncertainties on the background estimates include those on
\dyee, \dymm, and non-\WoZ\ leptons, which are estimated from data, as described
in Sections~\ref{sec:dydata} and~\ref{sec:fakesdata}. The uncertainties
on the remaining backgrounds are estimated through simulation.

The uncertainties on the single top, \VV, and \dytt\ backgrounds
arise from the same sources as for the \ttbar\ signal.
Uncertainties due to detector effects, described in Sections~\ref{sec:leptonSyst} and~\ref{sec:jesSyst}, 
contribute 10\% and are dominated by the energy scale uncertainty.
In events required to have at least one b tag, the uncertainty from \Bot\ tagging is roughly
$25\%$ for diboson and \dytt, and less than 10\% for single top events.
In addition, there is an uncertainty on each of the background production cross sections of 30\%.
This uncertainty is conservative with respect
to the uncertainties on the inclusive production rate, and is expected to cover
the uncertainties on the rate of these backgrounds in the phase space of the event selections
used in this analysis.
Measurements of the inclusive production rates for \PW\PW\ production~\cite{wwCMS2010} 
(the dominant among the contributions to \VV\ production in the SM) and  \dytt~\cite{dyttCMS2010}
are in good agreement with the SM.

\subsection{Cross section measurements per decay channel}
\label{sec:results_channels}
\label{sec:yields}
\begin{figure}[h!]
\begin{center}
  \includegraphics[width=0.47\textwidth]{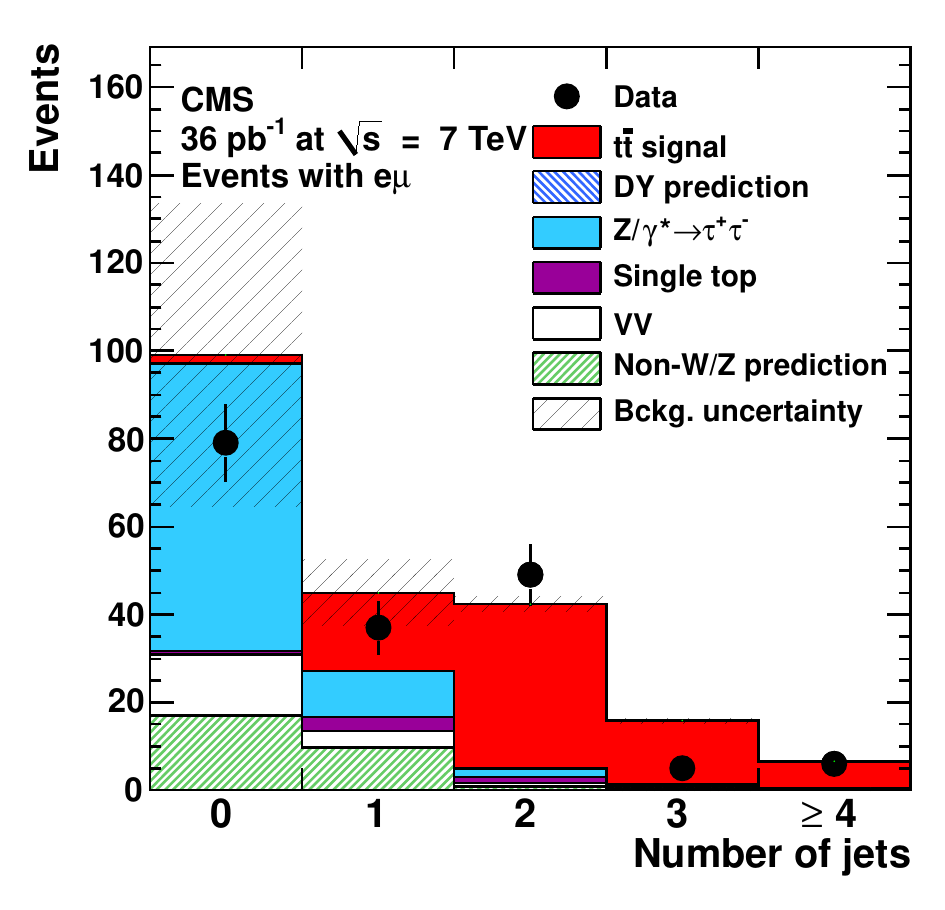}
  \includegraphics[width=0.47\textwidth]{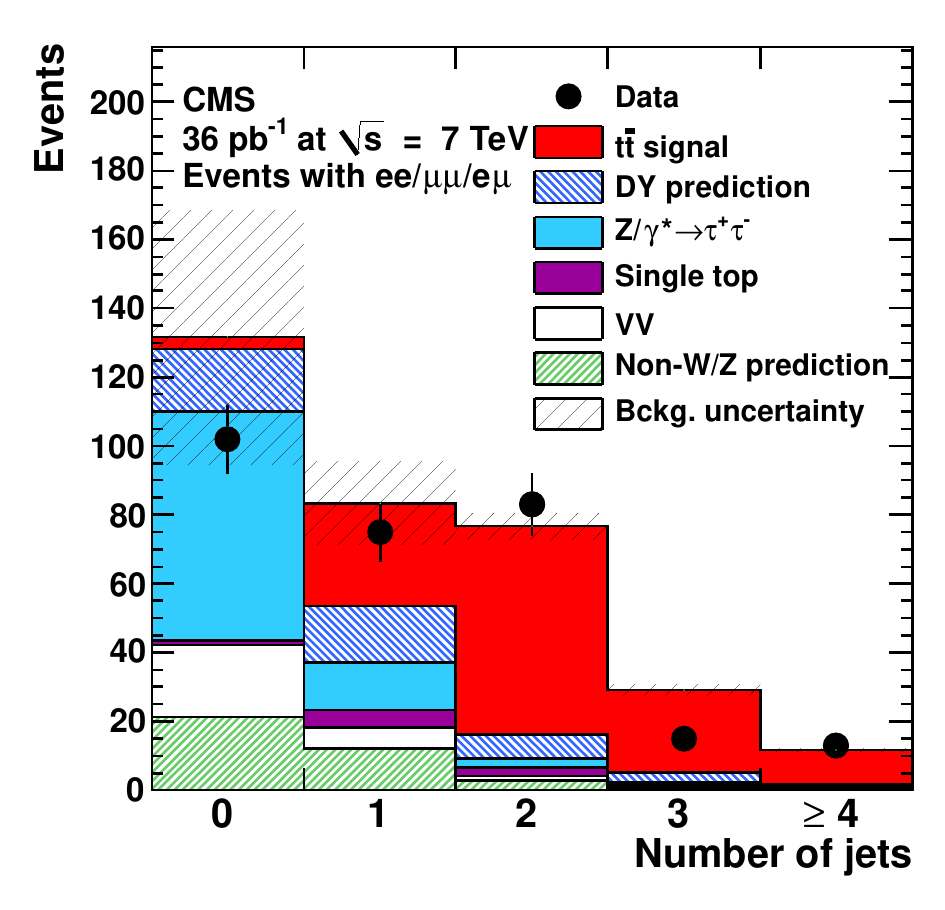}
\caption{Number of events passing the full dilepton selection criteria without a \Bot\ tag (points),
as a function of the jet multiplicity for \empm\ (left) and all dileptons (right).
There is no \met\ requirement for the \empm, and a requirement of $\met>30\GeV$ 
for the \eepm\ and \mmpm.
The expected distributions for the \ttbar\ signal and the background sources are shown by the
histograms.
The Drell--Yan and non-\WoZ\ lepton backgrounds are estimated from data, 
while the other  backgrounds are from simulation. 
The total uncertainty on the background contribution is displayed by the hatched region.\label{fig:njetsDD}}
\end{center}
\end{figure}

\begin{figure}[h!]
\begin{center}
  \includegraphics[width=0.47\textwidth]{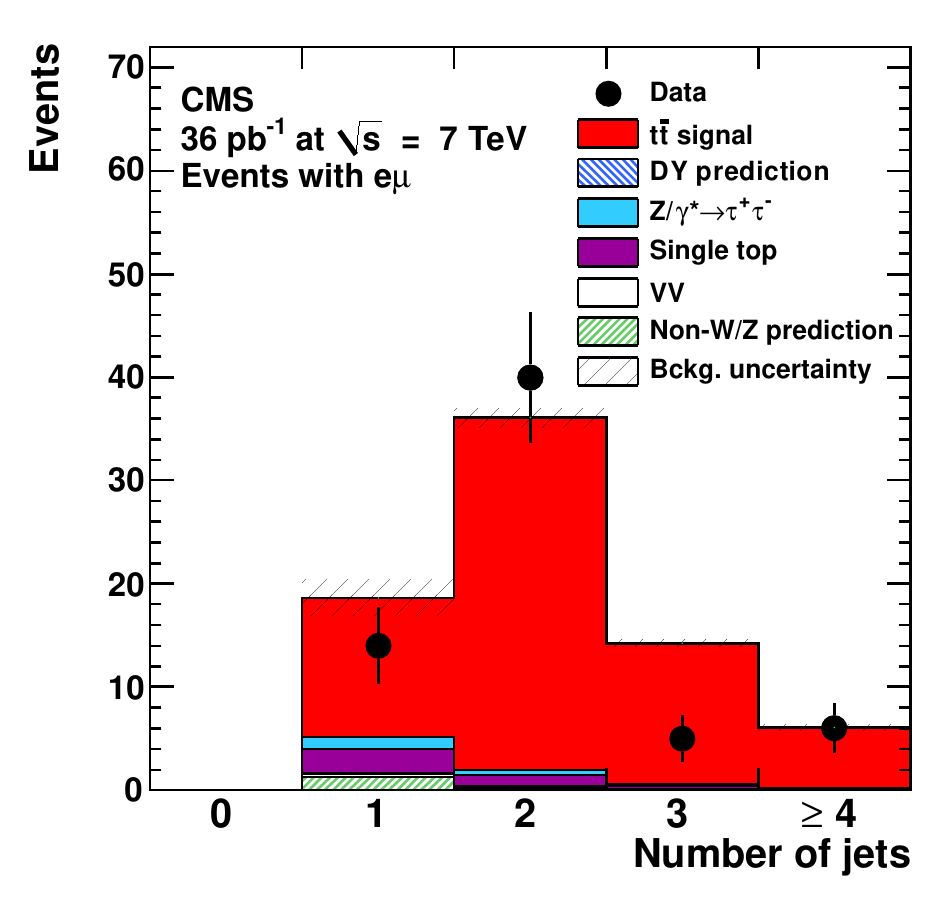}
  \includegraphics[width=0.47\textwidth]{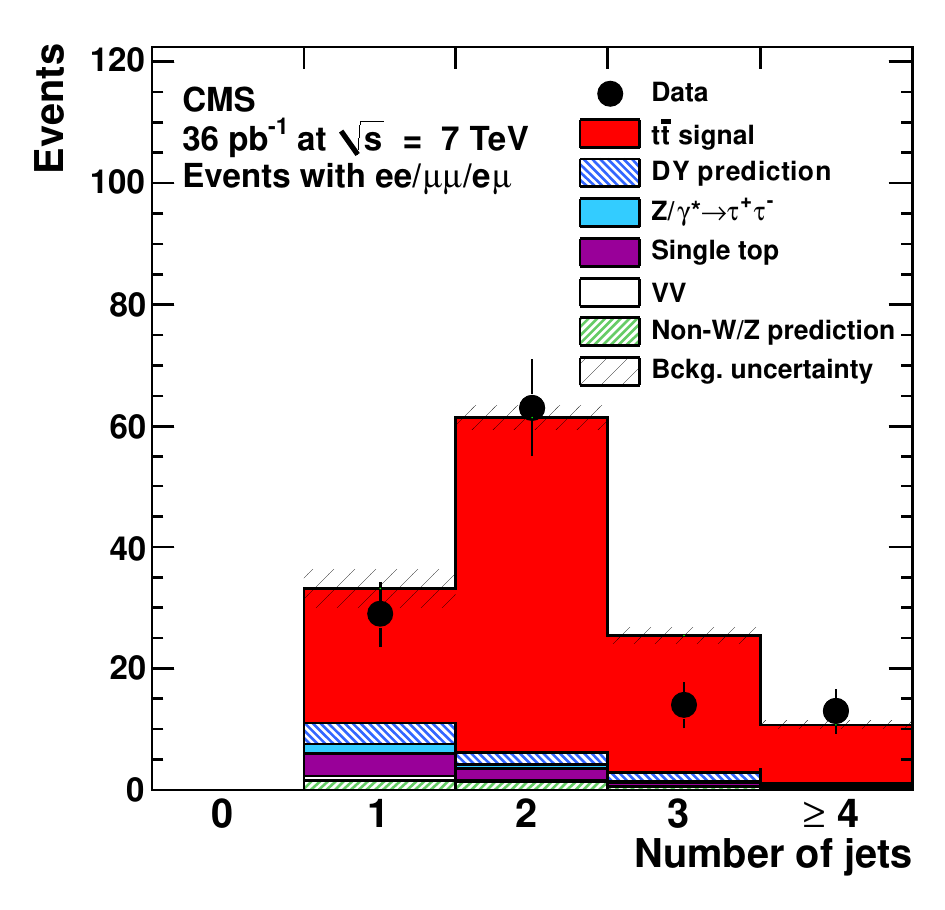}
\caption{Jet multiplicity for events passing
full  dilepton selection criteria with at least one b-tagged jet, otherwise the same as in Fig.~\ref{fig:njetsDD}.
\label{fig:njets1tagDD}}
\end{center}
\end{figure}

\begin{figure}[h!]
\begin{center}
  \includegraphics[width=0.47\textwidth]{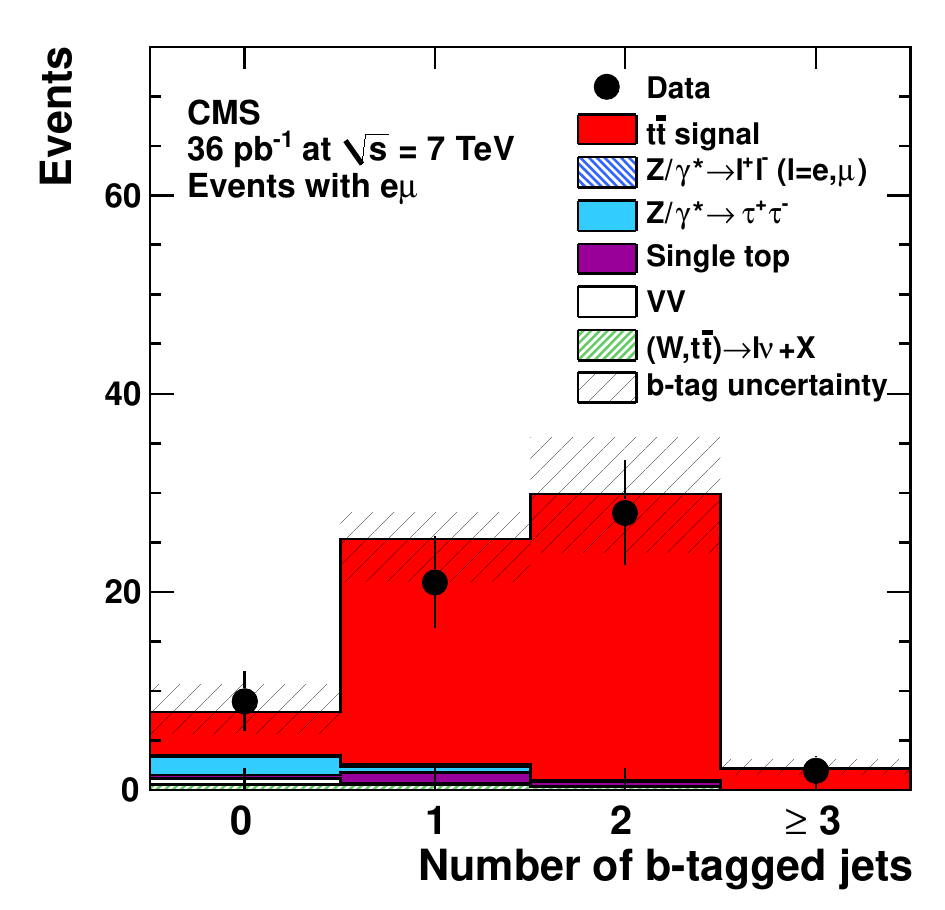}
  \includegraphics[width=0.47\textwidth]{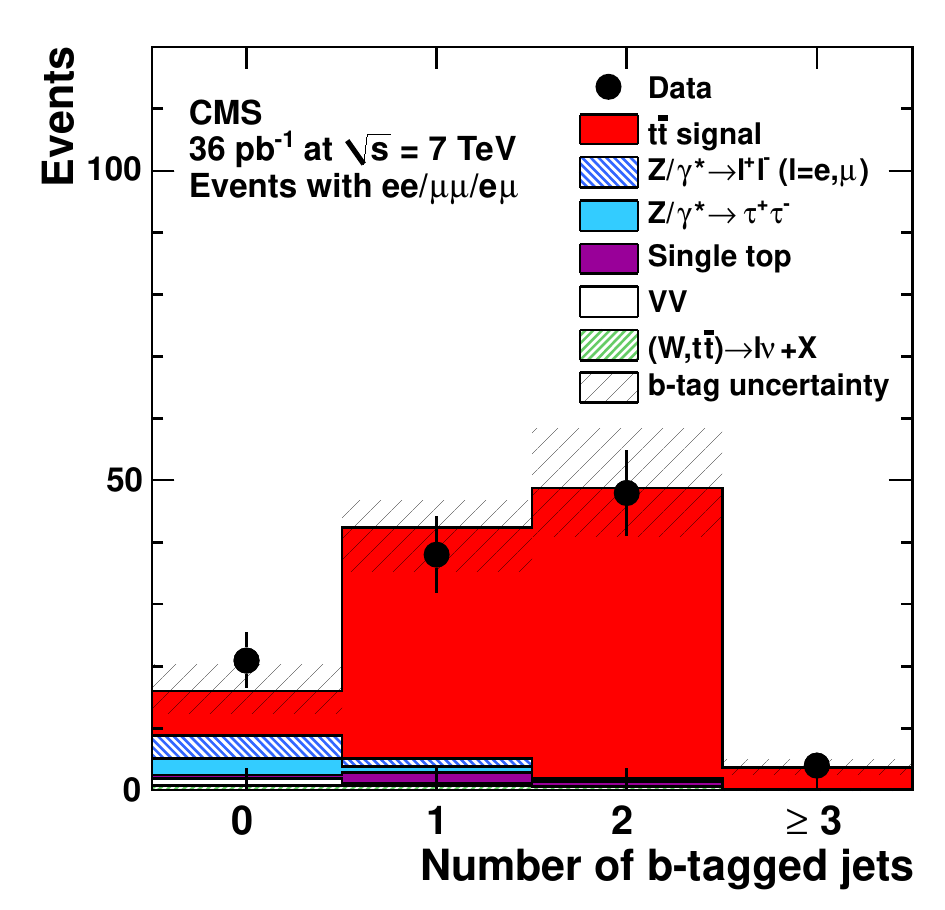}
\caption{Multiplicity of b-tagged jets in events passing
full  dilepton selection criteria with at least two jets 
compared to signal and background expectations from simulation.
The uncertainty on the number of signal events corresponding to the uncertainty in the selection of
b-tagged jets is displayed by the shaded area.
The distributions are for \empm\ (left) and all (right) final states combined.\label{fig:nbtags}}
\end{center}
\end{figure}

The expected numbers of signal and background dilepton events passing
all the selection criteria but without a \Bot\ tag are compared with data in Fig.~\ref{fig:njetsDD}
for \empm\ (left) and all (right), as a function of jet multiplicity.
There is a requirement of $\met>30\GeV$  for the \eepm\ and \mmpm
and no \met\ requirement for the \empm, as otherwise used for the signal selection 
of events with at least two jets.
Similar plots for events with at least one \Bot\ tag are shown in Fig.~\ref{fig:njets1tagDD}.
The observed numbers of events with zero or one jet can be used
as checks on the background predictions, since the main signal contribution
is for events with two or more jets.
The multiplicity of b-tagged jets observed in data is compared to the simulation in Fig.~\ref{fig:nbtags}.
Good agreement is found between the expected and observed  numbers of events in all channels.
A summary of the expected number of background events is compared with
the number of events observed in data in Table~\ref{tab:xsecChannels} for the channels used in the measurement.

The \ttbar\ production cross section is measured using:
\begin{equation}
\sigma(\pp\to\ttbar) = \frac{N - B}{\mathcal{A}  {L}},
\label{eqn:xsecA}
\end{equation}
where $N$ is the number of observed events; $B$ is the number of estimated background events;
$\mathcal{A}$ is the total acceptance relative to all produced \ttbar\ events,
 including the branching ratio to leptons,
the geometric acceptance, and the event selection efficiency
already corrected for differences between data and simulation;
and ${L}$ is the integrated luminosity.

Results of the signal and background estimates and events observed in data
in each of the three dilepton final states in events passing selections with at least two jets prior to and after
the  b-tagging requirement, and events with one jet are
summarised in Table~\ref{tab:xsecChannels}.
These nine sets of inputs are treated as separate measurements of the inclusive \ttbar\ production
cross section.
The uncertainties are propagated following Eq.~(\ref{eqn:xsecA}) for each selection in the following way:
the statistical uncertainty is given by ${\sqrt{N}}/{(\mathcal{A}  {L})}$;
the systematic uncertainty combines in quadrature the uncertainties on the backgrounds
and $\mathcal{A}$, where the relative uncertainties on $\mathcal{A}$ are reported in Table~\ref{tab:systSumm} as subtotal values;
the uncertainty on the luminosity (not reported in Table~\ref{tab:xsecChannels}) is 4\%, 
the same for all channels.
Consistent \ttbar\ cross section results are seen between the 9 measurements, 
within their relevant uncertainties.
The cross section measured in the \eepm\ and \mmpm\ final states 
with at least two jets and at least one \Bot-tagged
jet is more precise than the corresponding measurements in the same jet multiplicity without a \Bot-tagging
requirement, which results in a significant suppression of the backgrounds.
The situation is different in the \empm\ final state, where the b-tagging requirement
gives a slightly worse precision, primarily due to added uncertainty on the rate of \Bot-tagged events.
The measurements in events selected with one jet,
where the total number of events is smaller and the fraction of backgrounds is larger,
 have a substantially larger uncertainty compared to the
selections with at least two jets.

\begin{table}[h!]
\begin{center}
\caption{\label{tab:xsecChannels}
The number of dilepton events observed in data, the background estimates, 
the total signal acceptance $\mathcal{A}$ (with systematic uncertainties), 
and the resulting \ttbar\ cross section measurements are shown for each of the dilepton samples,
from samples of events with one and more than one jet, and with and without at least one \Bot\ tag.
The simulated background estimates are the sum of the \dytt, \VV, and single top contributions.
The two uncertainties on the cross section measurements are the statistical and systematic contributions, 
respectively, excluding the $4\%$ luminosity normalization uncertainty.
}
\begin{tabular}{lccc}\hline\hline
Final state 		& \eepm			& \mmpm			& \empm			\\ \hline
\multicolumn{4}{c}{At least two jets, no b-tagging requirement}						\\ \hline
Events in data		& 23			& 28			& 60			\\ \hline
Simulated backgrounds	& $1.4\pm 0.3$		& $1.5\pm 0.3$		& $5.2\pm 1.2$		\\
\dyee/\mmpm		& $3.0\pm 1.8$		& $7.4\pm4.1$		& --			\\
Non-\WoZ\		& $1.1\pm 1.4$		& $0.6\pm1.1$		& $1.4\pm1.6$		\\ \hline
All backgrounds		& $5.5\pm 2.3$		& $9.5\pm4.3$		& $6.7\pm2.0$		\\
Total acceptance $\mathcal{A}$ (\%)	& $0.259\pm0.021$	& $0.324\pm0.025$	& $0.928\pm0.057$	\\
Cross section (pb)	& $189\pm52\pm29$	& $159\pm45\pm39$	& $160\pm23\pm12$	\\ \hline
\multicolumn{4}{c}{At least two jets, at least one \Bot-jet}					\\ \hline
Events in data		& 15			& 24			& 51			\\ \hline
Simulated backgrounds	& $0.7\pm0.2$		& $0.8\pm0.3$		& $2.5\pm0.7$		\\
\dyee/\mmpm		& $0.7\pm0.7$		& $2.6\pm1.8$		& --			\\
Non-\WoZ\		& $0.9\pm1.2$		& $0.3\pm0.8$		& $0.5\pm1.1$		\\ \hline
All backgrounds		& $2.3\pm 1.4$		& $3.8\pm2.0$		& $3.0\pm1.4$		\\
Total acceptance $\mathcal{A}$ (\%)	& $0.236\pm0.022$	& $0.303\pm0.028$	& $0.857\pm0.068$	\\
Cross section (pb)	& $150\pm46\pm22$	& $186\pm45\pm25$	& $156\pm23\pm13$	\\ \hline
\multicolumn{4}{c}{One jet, no b-tagging requirement}						\\ \hline
Events in data		& 8			& 10			& 18			\\ \hline
Simulated backgrounds	& $1.6\pm0.4$		& $1.9\pm0.4$		& $3.6\pm0.9$		\\
\dyee/\mmpm		& $0.2\pm0.3$		& $5.2\pm4.3$		& --			\\
Non-\WoZ\		& $0.3\pm0.5$		& $0.1\pm0.4$		& $1.3\pm1.3$		\\ \hline
All backgrounds		& $2.1\pm 0.7$		& $7.1\pm4.3$		& $4.9\pm1.5$		\\
Total acceptance $\mathcal{A}$ (\%)	& $0.058\pm0.007$	& $0.074\pm0.008$	& $0.183\pm0.024$	\\
Cross section (pb)	& $282\pm135\pm45$	& $107\pm119\pm163$	& $200\pm65\pm35$	\\ \hline \hline
\end{tabular}
\end{center}
\end{table}

In addition to the selections used for the main results presented in this analysis,
alternative selections were applied to the same data sample
and most of the steps of this analysis were reproduced.
One analysis used calorimeter jets and missing transverse energy, both corrected using
tracks reconstructed in the silicon tracker~\cite{METPAS2,JETPAS}.
Another analysis applied lepton identification and isolation requirements
based on quantities provided by the particle flow algorithm~\cite{METPAS2,PFTPAS10}.
Corresponding analyses based on these alternative selections
provide results compatible with the performance of the analysis presented here.

\subsection{Combination of cross section measurements}
\label{sec:xs_combined}
The cross section measurements detailed in the previous section are combined
to produce a final overall value.
The combination is done using the best linear unbiased
estimator (BLUE) technique~\cite{Lyons1988110},
which accounts for correlations between different contributions.
This combination includes statistically correlated contributions from the
events selected with at least two jets with and without a b-tagging requirement.
The correlation coefficients estimated with toy simulation are
 75\%, 85\%, and 90\% in the \eepm, \mmpm, and \empm\ final states, respectively.
The combination of all nine measurements shown in Table~\ref{tab:xsecChannels} 
has a $\chi^2$ value of  $2.5$ for eight degrees of freedom.
The combined value of the cross section is
\begin{eqnarray}
\label{eq:xsecAllLum}
\sigma({\pp\to\ttbar}) = 168 \pm 18 \,({\rm stat.})\pm 14 \,({\rm syst.})\pm 7 \,({\rm lumi.}){\rm~pb}.
\end{eqnarray}

Alternatively, a combination of
statistically independent measurements was performed using
non-overlapping contributions: events with only one jet and events with
dielectrons and dimuons with at least two jets and at least one b-tagged jet
are combined with electron-muon events with at least two jets.
A result consistent with the value in Eq.~(\ref{eq:xsecAllLum}) was obtained in this combination.

\subsection{Ratio of \ttbar\ and \dy\ cross sections}
\label{sec:sample_normalization}
A measurement of the ratio of the \ttbar\ and \dy\ production cross sections 
is less sensitive to the various systematic uncertainties than the \ttbar\ cross section itself.
The ratio does not depend on the integrated luminosity
and has a substantially reduced dependence on the lepton selection efficiencies.
Events from \dyee\ and \mmpm\ selected 
by requiring just two identified, oppositely charged  isolated leptons,
as described in Section~\ref{sec:event_selection}, are used to measure the \dy\ production cross section.
Since the same lepton selection criteria are used,
the simulation-to-data corrections on the lepton efficiencies
cancel out in the ratio.

The number of data events passing the event selection criteria with a dilepton
invariant mass in the range of $76<\mll<106\GeVcc$
 is 10\,703 (13\,594) for the \eepm\ (\mmpm) final state.
Backgrounds are less than 1\% and are ignored.
After correcting for the lepton selection efficiency described in Section~\ref{sec:leptonSyst}
using the NLO generator \POWHEG, the measured
production cross section averaged for \dyee\ and \mmpm\ is  $961\pm 6$~pb, where the uncertainty
is statistical.
The \dy\ cross section reported here is computed relative to the dilepton final states 
in the range of $60<\mll<120\GeVcc$, as reported in~\cite{inclusWZ3pb,wzPAS2010}.
These can be compared to the expected value of $972\pm 42$~pb, computed at NNLO with {\sc fewz}.
There is a remaining $2.2\%$  systematic uncertainty on the average \dy\ cross section 
measurement that is relevant  for the ratio:
$2.0\%$ for the \mmpm\ and $2.5\%$ for the \eepm events, of which $2.0\%$ is common.

The resulting ratio of the \ttbar\ and \dyee\ and \mmpm\ cross sections is found to be:
\begin{eqnarray}
\label{eq:xsecRatioDY}
\frac{\sigma({\pp\to\ttbar})}{\sigma({\pp\to\dy\to\eepm/\mmpm})} 
	= 0.175 \pm 0.018 \,({\rm stat.})\pm 0.015 \,({\rm syst.}).
\end{eqnarray}
The relative total uncertainty of 14\% on the ratio is marginally better than the total uncertainty on
the \ttbar\ cross section, as the dominant uncertainties specific to the \ttbar\ measurement remain
and the \dy\ part of the measurement introduces an additional small uncertainty.
The total uncertainty on the ratio is approximately the same as that on the ratio of the SM
predictions for the cross sections.
Thus, this measurement can already be useful in restricting the parameters (e.g., PDFs) 
used in the SM predictions.

\section{Measurement of the top quark mass}
\label{sec:mass}
Many methods have been developed for measuring the top quark mass $m_{\rm top}$ in the dilepton channel.  
The Matrix Weighting Technique (MWT)~\cite{D01998} was one approach used in the first measurements with this channel~\cite{D01998,CDF1997}.
Other approaches were developed later, for example the fully kinematic method (KIN)~\cite{kin}. The average of the measurements in the dilepton 
channel is $171.1\pm2.5$~GeV\!/\!$c^2$~\cite{topmass_tev}.
In the present measurement, improved versions of the MWT and KIN algorithms are used. 
The improved methods KINb (KIN using \Bot-tagging) and AMWT (analytical MWT) are discussed in the following in detail.

The reconstruction of $m_{\rm top}$ from dilepton events leads to an under-constrained system, since the dilepton channel contains at least two neutrinos in the final state.
For each \ttbar\ event, the kinematic properties are fully specified by 24 variables, which are the four-momenta of the 6 particles in the final state.
Of the 24 free parameters, 23 are known from different sources: 14 are measured (the three-momenta of the jets and leptons, and the two components of the \met) and 9 are constrained.
The system can be constrained by imposing the \PW\ boson mass to its measured value (2 constraints), by setting the top and anti-top quark masses to be the same (1), 
and the masses of the 6 final state particles to the values used in the simulation~\cite{madgraph} (6).
This still leaves one free parameter that must be constrained by using some hypothesis that depends on the method employed.

A subset of the events selected for measuring the top quark pair production cross section is used to determine $m_{\rm top}$.
To ensure a good kinematic reconstruction, only events with at least two jets are used.
In addition to the $\met > 30\GeV$ requirement for the \eepm\ and \mmpm\ dilepton events, a $\met > 20\GeV$ cut is introduced 
for the \empm\  channel in order to achieve a better $\vec{\met}$ direction resolution, which directly reflects on the $m_{\rm top}$ resolution.
 
A key difference with respect to previous measurements of $m_{\rm top}$ is the choice of the jets used to reconstruct the top quark candidates. 
Because of initial-state radiation, 
the two leading jets (\ie, the jets with the highest \pt) 
may not be the ones that originate from the decays of the top quarks. 
The fraction of correctly assigned jets can be increased by using the information provided by \Bot-tagging. 
Therefore, \Bot-tagged jets in an event are used in the reconstruction, even if they are not the leading jets.
When no jet is \Bot-tagged, the two leading jets are used. If there is a single \Bot-tagged jet in the event, it is supplemented by the leading untagged jet.
Using MC simulation, we find that the fraction of events in which the jets used for the reconstruction are correctly matched to the partons from the top quark 
decay is significantly increased by this method.
The number of observed and expected events in each \Bot-tag multiplicity is shown in Table~\ref{tab:dilyield}.

\begin{table}[htp]
\begin{center}
\caption{Total number of dilepton events in each \Bot-tag multiplicity.
The quoted uncertainties include the statistical uncertainty and the uncertainties for jet energy scale variation and the \Bot/mis-tagging efficiency variation,
which cancels out in the last row.
The uncertainty due to the luminosity is not shown.
}
\label{tab:dilyield}
\begin{tabular}{l|c|c|cc}\hline 
\Bot-tag multiplicity          & Data            & Total expected                 & \ttbar\ signal                 & Total background \\\hline\hline
$=$ 0 \Bot-tag		      & 19              &  15.7 $\pm$ 0.6 $^{+12}_{-8}$  & 6.8 $\pm$ 0.2  $^{+7}_{-3}$   & 8.9 $\pm$ 0.6 $^{+6}_{-5}$ \\
$=$ 1 \Bot-tag                 & 35              &  40.6 $\pm$ 0.5 $^{+17}_{-13}$ & 35.5 $\pm$ 0.4 $^{+9}_{-8}$   & 5.1 $\pm$ 0.4 $^{+8}_{-6}$ \\
$\geq$ 2 \Bot-tags	      & 48              &  51.4 $\pm$ 0.5 $^{+14}_{-16}$ & 49.2 $\pm$ 0.5 $^{+11}_{-15}$ & 2.2 $\pm$ 0.2 $^{+3}_{-1}$ \\\hline
Total                         & 102             &  107.7 $\pm$ 0.9 $^{+3}_{-2}$  & 91.5 $\pm$ 0.7 $^{+2}_{-1}$   & 16.2 $\pm$ 0.7 $^{+1}_{-1}$ \\\hline
\end{tabular}
\end{center}
\end{table}

\subsection{Mass measurement with the KINb method}
\label{sec:mass_reconstruction_kin}
\begin{figure}[htp]
\centering
\includegraphics[width=0.5\textwidth]{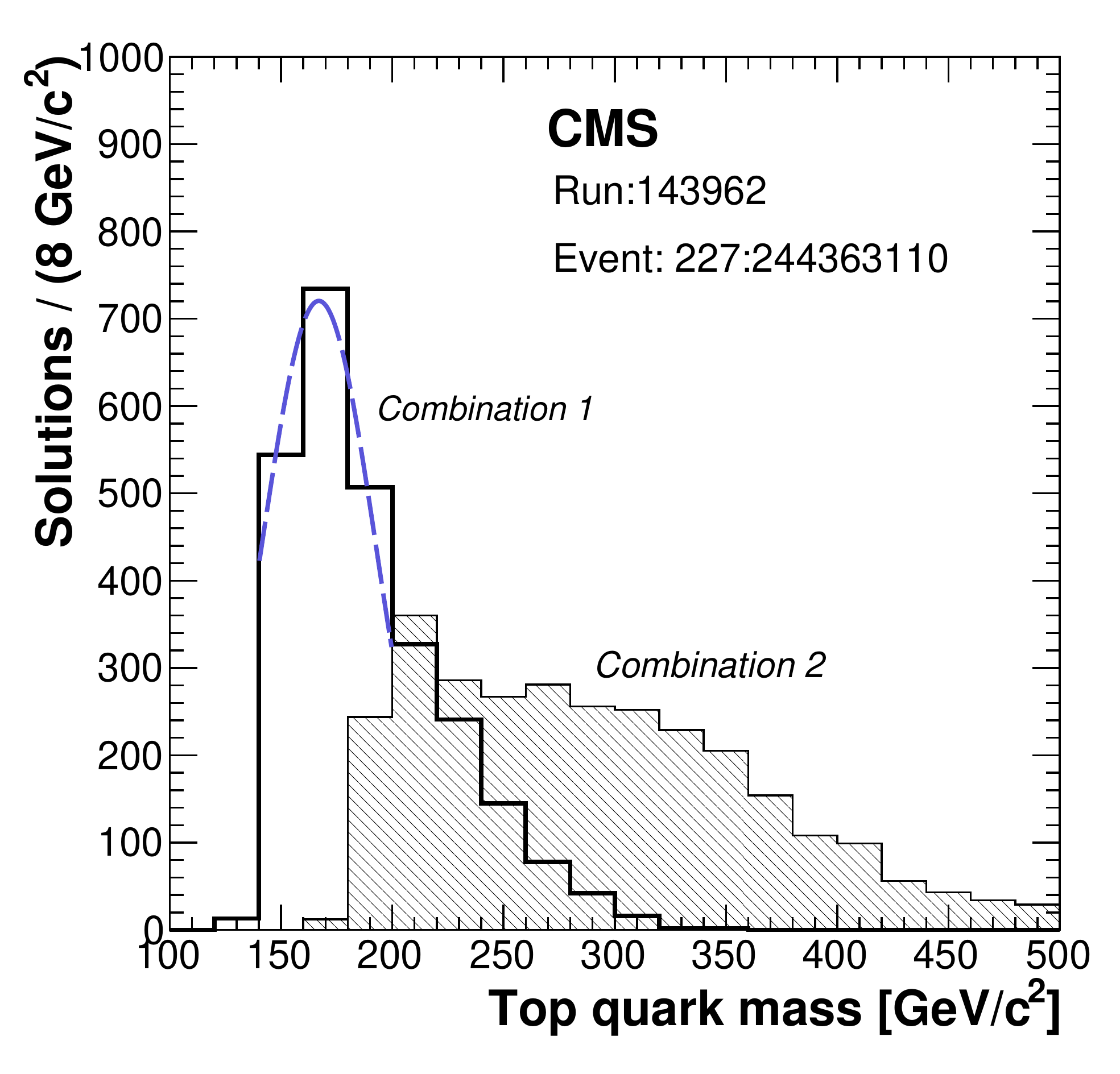}
\caption{Top quark mass solutions for the KINb method for the two lepton-jet combinations after smearing the jet energy resolution for one selected event in data. 
The combination \#1 is chosen in this case; the dashed line corresponds to the Gaussian fit used to estimate $m_{\rm KINb}$ (see text).}
\label{fig:masssolutions} 
\end{figure}

In the fully kinematic method KINb, the kinematic equations describing the \ttbar\ system are solved many times per event for each lepton-jet combination. 
Each time, the event is reconstructed by varying independently the jet $\pt$, $\eta$ and $\phi$, and the $\vec{\met}$ direction;
resolution effects are accounted for by reconstructing the event 10000 times, each time drawing random numbers from 
a normal distribution with mean equal to the measured values and width equal to the detector resolution obtained from the data.
For each variation, the unmeasured longitudinal momentum of the \ttbar\ system $p_z^{\rm t\bar t}$ is also drawn randomly from a simulated distribution.
The $p_z^{\rm t\bar t}$ value, which is minimally dependent on $m_{\rm top}$, is used to fully constrain the \ttbar\ system.
For each set of variations and each lepton-jet combination,
the kinematic equations can have up to four solutions, and the one with the lowest invariant mass of the \ttbar\ system is accepted
if the difference between the two top quark masses is less than $3\GeVcc$. 
For each event, the accepted solutions of the kinematic equations corresponding to the two possible lepton-jet combinations are counted.
The combination with the largest number of solutions is chosen, and the mass value $m_{\rm KINb}$ 
is found by fitting the $m_{\rm top}$ distribution of all the solutions from the event with a Gaussian function in a $50\GeVcc$ window around the peak of the distribution.
When the number of solutions found for the two combinations is similar (\ie, with a difference of less than 10\%), 
the combination with the highest peak is chosen.
An example of the distributions from the two lepton-jet combinations for one event is shown in Fig.~\ref{fig:masssolutions}.
Events with no solutions do not contribute to the $m_{\rm top}$ measurement; in simulation, solutions are found for 98\% of signal events and 80\% of background events, 
thereby providing additional background rejection. The lepton-jet pair is correctly assigned in 75\% of the cases.

\begin{figure}[htp]
\centering
\includegraphics[width=0.5\textwidth]{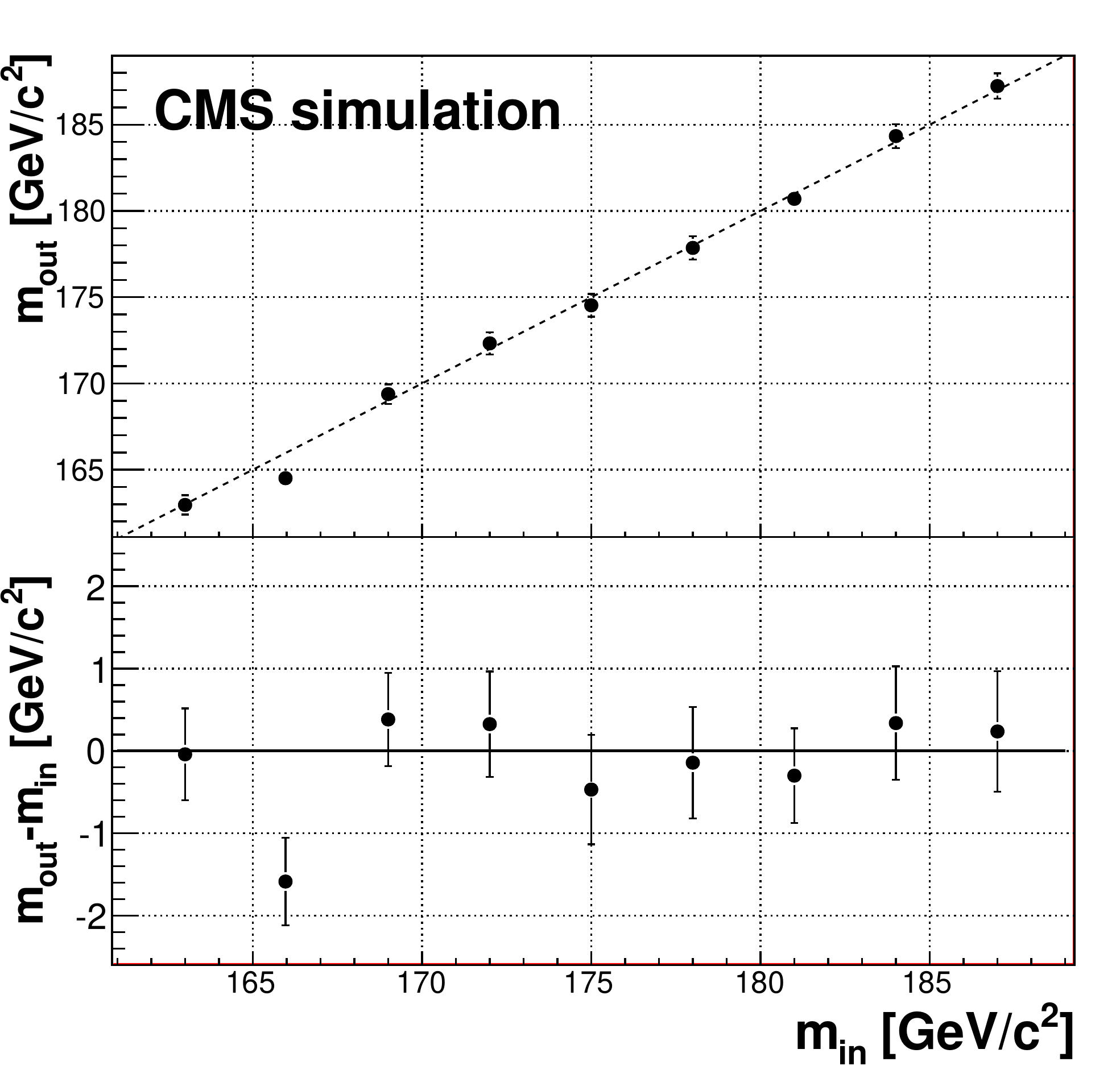}
\caption{
(Top) Fitted top quark mass values $m_{\rm out}$ using the KINb algorithm from simulated pseudo-experiments, including signal and background processes, 
as a function of the actual top quark mass used in the simulation ($m_{\rm in}$). A linear fit to the points is also shown.
(Bottom) The difference (bias) between the linear fit and the actual reconstructed values from the pseudo-experiments.
The bias is shown after calibrating the signal parametrisation.}
\label{fig:signalfitcalib} 
\end{figure}

Because of the presence of background and misreconstructed signal, a two-component (signal plus background), unbinned maximum likelihood fit of the $m_{\rm KINb}$ 
distribution is used to obtain an unbiased estimate of $m_{\rm top}$. 
The free parameters in the likelihood fit are $m_{\rm top}$ and the numbers of signal and background events.
The fit uses signal and background shapes of the $m_{\rm top}$ distribution that are produced from simulation for different values of $m_{\rm top}$, and which are fixed in the fit.
The signal and background shapes may resemble each other as a function of $m_{\rm top}$.
Therefore, the number of background events is constrained to the expected value by a Gaussian term in the likelihood.
The signal shape is obtained with a simultaneous fit to
simulated \ttbar\ samples, generated with $m_{\rm top}$ values between 151 and $199\GeVcc$ in steps of $3\GeVcc$, of a Gaussian+Landau distribution 
with parameters that are linear functions of $m_{\rm top}$.  
Separate distributions are used for the three samples with 0, 1, and 2 or more \Bot-tagged jets, and the backgrounds are added in the expected proportions.
The relative contribution of \PZ+jet events to the total background is determined from data
by counting the number of dilepton events 
with an invariant mass near the \PZ-boson peak ($|m_{\ell\ell} - m_Z|<15\GeVcc$).
The other background contributions are taken from simulation.

\begin{figure}[htp]
\includegraphics[width=0.48\textwidth]{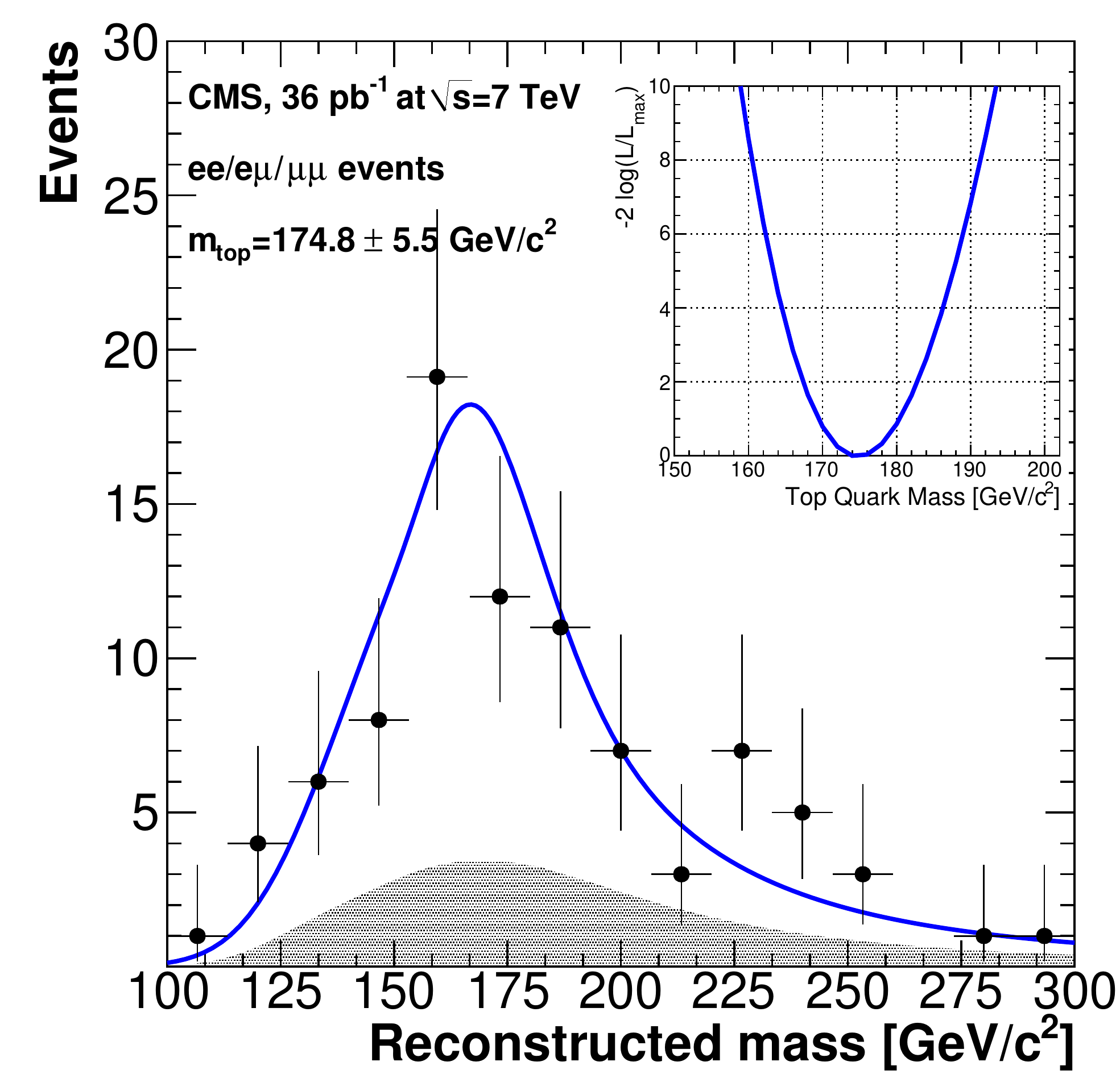} \hfill
\includegraphics[width=0.48\textwidth]{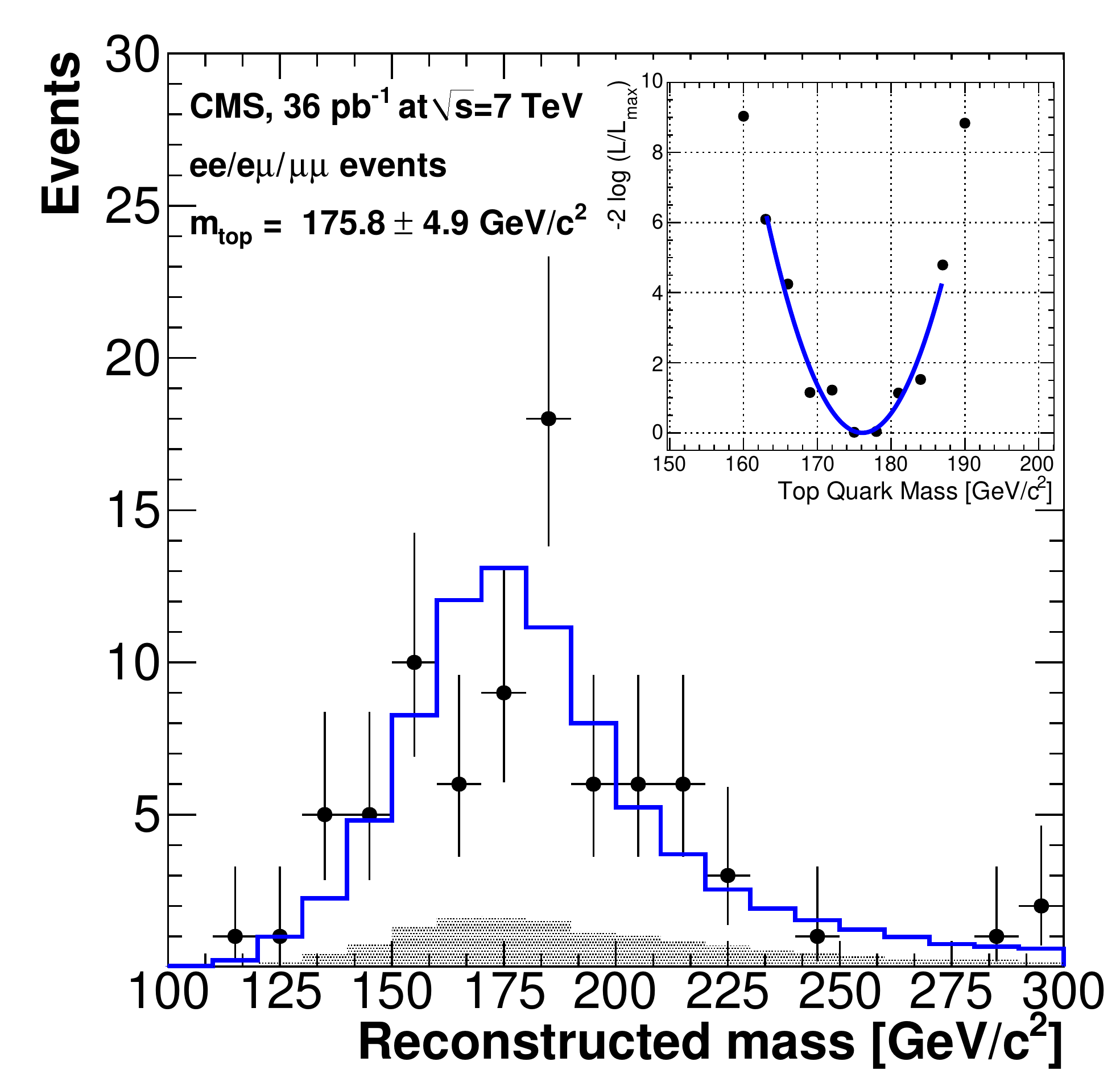}
\caption{Reconstructed top quark mass distributions from the KINb (left) and AMWT (right) methods. Also shown are the total background
plus signal models, and the background-only shapes (shaded). The insets show the likelihoods as functions of $m_{\rm top}$.}
\label{fig:topmass_fit}
\end{figure}

In order to minimise 
any residual bias resulting from the parameterisations of the signal and background $m_{\rm KINb}$ distributions, pseudo-experiments are performed using simulated
dilepton events generated with different $m_{\rm top}$ values.
The resulting $m_{\rm top}$ distributions are used to calibrate the parametrisation of the signal template.  We find an
average bias on $m_{\rm top}$ of $-0.7\pm 0.2\GeVcc$, which we use to correct our final value.
Figure~\ref{fig:signalfitcalib} shows the linearity (top plot) and the residual bias (bottom plot) of the fit, after applying the calibration corrections.
The left plot in Fig.~\ref{fig:topmass_fit} shows the $m_{\rm KINb}$ mass distribution from data and the result of the fit. The insert displays
the variation of  the likelihood ${\cal L}$ used in the fit, $-2 \ln({\cal L}/{\cal L}_{\rm max})$ as a function of $m_{\rm top}$.

\subsection{Mass measurement with the AMWT method}
\label{sec:mass_reconstruction_mwt}
In the analytical matrix weighting technique (AMWT), the kinematic equations describing the \ttbar\ system are solved many times per event. 
The mass of the top quark is used to fully constrain the \ttbar\ system.
The analytical method proposed in Ref.~\cite{PhysRevD.73.054015, *PhysRevD.78.079902} is used to determine the momenta of the two neutrinos. 
For a given top quark mass hypothesis, 
the constraints and the measured observables restrict the transverse momenta of the neutrinos to lie on ellipses
in the $p_x$-$p_y$ plane. 
If we assume that the measured missing transverse energy is solely due to the neutrinos, 
the two ellipses constraining the  transverse momenta of the neutrinos can be obtained, and
the intersections of the ellipses provide the solutions that fulfil the constraints.
With two possible lepton-jet combinations, there are up to eight solutions for the neutrino momenta for a given hypothesis of the top quark mass.

Each event is reconstructed many times using a series of input $m_{\rm top}$ values between 100 and $300\GeVcc$ in $1\GeVcc$ steps.
Typically, solutions are found for the neutrino momenta that are consistent with all constraints for large intervals of $m_{\rm top}$.
In order to determine a preferred value of $m_{\rm top}$, the following weight is assigned to each solution~\cite{DG}:

\begin{equation}
w = \left\{\sum F(x_1)F({x_2})\right\}p(E_{\ell^+}^*|m_{\rm top})p(E_{\ell^-}^*|m_{\rm top}),
\end{equation}

where $x_i$ are the Bj\"orken $x$ values of the initial-state partons, $F(x)$ is the PDF, the summation is over the possible leading-order initial-state partons 
($\mathrm{u\bar{u}}$, $\mathrm{\bar{u}u}$, $\mathrm{d\bar{d}}$, $\mathrm{\bar{d}d}$, and $\mathrm{gg}$), and the term $p(E^*|m_{\rm top})$ is the probability of observing a charged lepton 
of energy $E^*$ in the rest frame of the top quark, for a given $m_{\rm top}$.
For each value of $m_{\rm top}$, the weights $w$ are added for all solutions. Detector resolution effects are accounted for by 
reconstructing the event 1000 times, each time drawing random numbers for the jet momenta from a normal distribution with mean equal to the measured momentum and width 
equal to the detector resolution. The weight is averaged over all resolution samples for each event and $m_{\rm top}$ hypothesis.
For each event, the $m_{\rm top}$ hypothesis with the maximum averaged weight is taken as the reconstructed top quark mass $m_{\rm AMWT}$.
Events that have no solutions or that have a maximum weight below a threshold value are discarded. 
Based on simulations, we expect this requirement to remove about 9\% of the \ttbar\ and 20\% of the \PZ+jet events from the sample.

A likelihood $\cal L$ is computed for values of $m_{\rm top}$ between 151 and 199\GeVcc
in steps of $3\GeVcc$, using data in the range $100<m_{\rm AMWT}<300\GeVcc$.
A unique shape determined from MC is used  for each \Bot-tag category, where the peak mass distribution of each individual contribution is added according to its expected relative contribution. For the \PZ+jet background, both the distribution and its relative contribution are derived from data in the \PZ-boson mass window (c.f.\ Section~\ref{sec:dydata}).
For the other contributions (signal, single top production, non-dileptonic decays of \ttbar\ pairs), the distributions predicted by the simulation are used. Further background contributions are
negligible and are not taken into account in the fit.

\begin{figure}[htbp]
  \begin{center}
    \includegraphics[width=5cm]{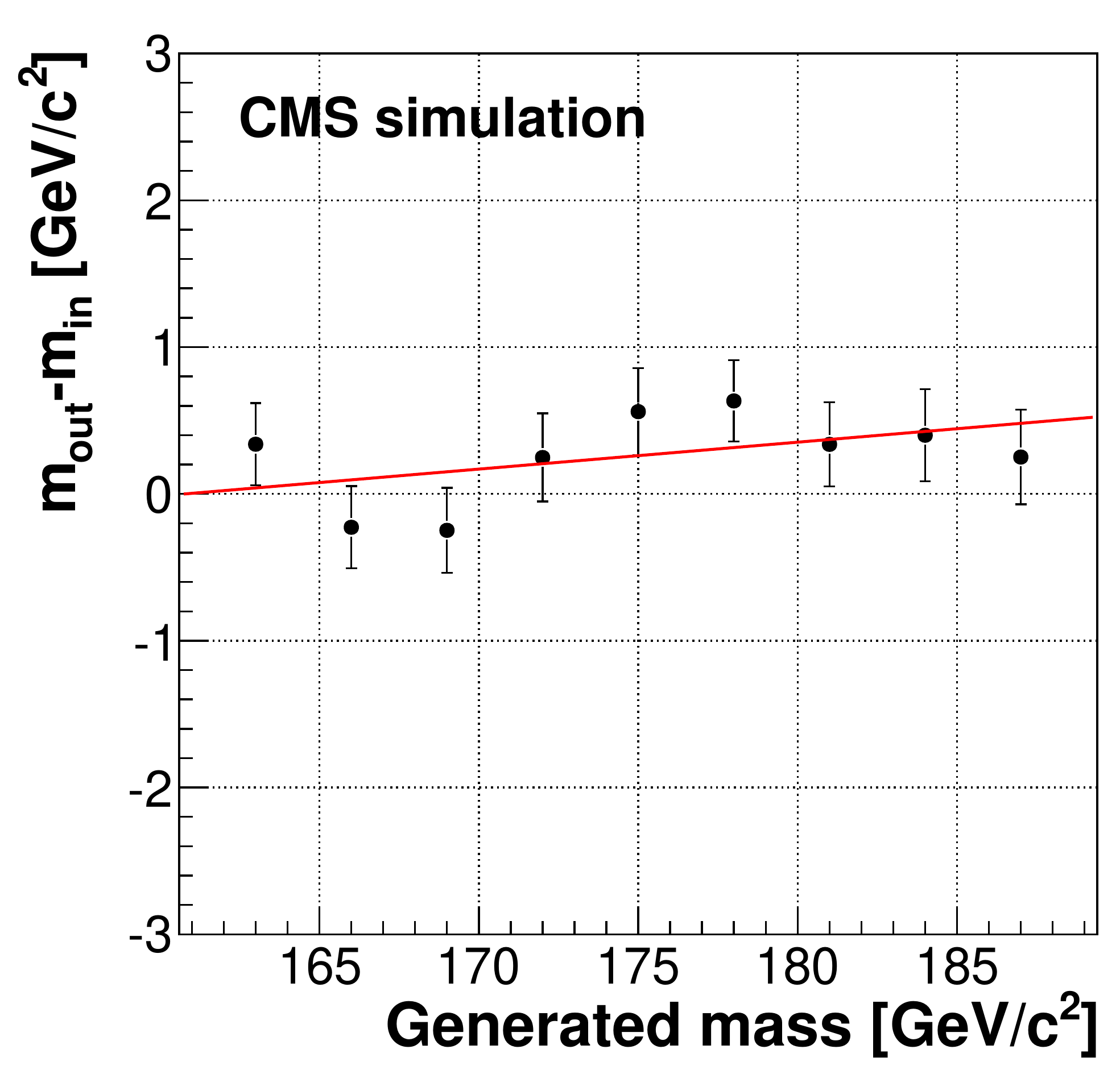}
    \includegraphics[width=5cm]{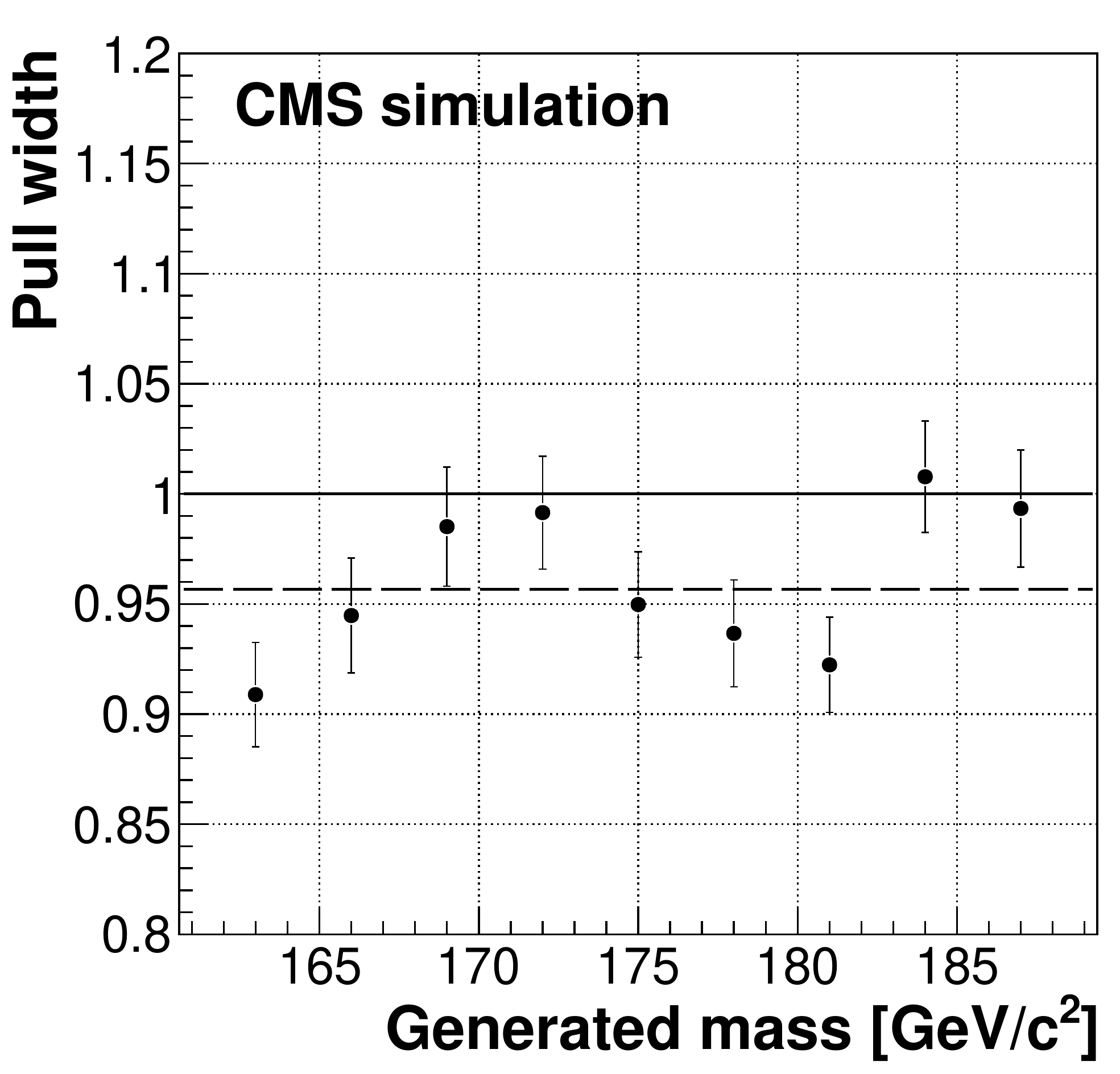}
    \caption{Mean mass bias (left) and pull width (right) for different mass hypotheses in pseudo-experiments for the AMWT method.
    The red solid line represents the linear fit used to determine the correction to apply in order to minimise the residual bias.
    The average pull width of 0.96 is shown with a dashed line. 
    }
    \label{fig:calib_B_tag}
  \end{center}
\end{figure}

We determine the bias of this estimate using ensembles of pseudo-experiments based on the expected numbers of signal and background events, as shown in Fig.~\ref{fig:calib_B_tag}. 
A small correction of $0.3\pm0.1\GeVcc$ is applied to the final result to compensate for the residual bias introduced by the fit (Fig.~\ref{fig:calib_B_tag}, left).
The width of the pull distribution is on average about 4\% smaller than 1.0, indicating that the statistical uncertainties are overestimated (Fig.~\ref{fig:calib_B_tag}, right). 
The statistical uncertainty of the measurement is therefore corrected down by 4\%.

Figure~\ref{fig:topmass_fit} (right) 
shows the predicted distribution of $m_{\rm AMWT}$ summed over the three \Bot-tag categories for the case of simulated $m_{\rm top}=175\GeVcc$, superimposed on the distribution observed in data.
The minimum of $-\ln(\cal L)$, determined from a fit to a quadratic function, is taken as the measurement of $m_{\rm top}$.
  
\subsection{Systematic uncertainties}
\label{sec:mass_systematics}
The sources of systematic uncertainties considered for the mass measurement are the same as those described in Section~\ref{sec:systematics}, 
and the most important contributions are summarised in Table~\ref{tab:syst}. 

The dominant source of uncertainty is the jet energy scale (JES), composed of an overall jet energy scale and a \Bot-jet specific energy scale~\cite{JESPAS}. 
The jet and lepton energy scales have a direct impact on the measurement
since they shift the momenta of the reconstructed objects, and hence the measured mass.
The JES yields the largest single uncertainty, and is estimated 
by generating pseudo-experiments from MC event samples for which the JES is varied by its uncertainty, and fitting them with the templates derived with the nominal JES.

The modelling of the underlying event is studied by comparing results from simulated
pseudo-experiments generated with \MADGRAPH and \PYTHIA using two different parameter sets for the generation 
of the underlying event (Z2 and D6T)~\cite{Field:2010bc}.
The uncertainty due to pileup is evaluated from pseudo-experiments containing \ttbar\ events with the inclusion of a number of pileup events similar to that in data
(approximately two pileup events on average).
An increase in the reconstructed mass is observed, and the full shift is used as the uncertainty.
The effect due to the scale used to match clustered jets to partons (\ie, jet-parton matching) is estimated with dedicated samples generated
by varying the nominal matching \pt\ thresholds by factors of 2 and 1/2.
Effects due to the definition of the renormalisation and factorisation scales 
used in the simulation of the signal are
studied with dedicated MC samples with the scales varied by a factor of two. 
The residual bias resulting from the fit calibration procedure is estimated from the deviation in the reconstruction of the top quark mass 
measured from pseudo-experiments using different mass points, as described in Sections~\ref{sec:mass_reconstruction_kin} and~\ref{sec:mass_reconstruction_mwt}.

Additional uncertainties come from the modelling of the signal templates (MC generator), which are studied by comparing the results of the pseudo-experiments 
using the reference samples to samples from the \ALPGEN and \POWHEG generators.
The uncertainties related to the PDF used to model the hard scattering of the proton-proton collisions is estimated by using pseudo-experiments for which
the distribution of $m_{\rm top}$ is obtained after varying the PDF by its uncertainties using the PDF4LHC prescription~\cite{cteq66,pdf4lhcInterim}.
The uncertainty due to \Bot-tagging is evaluated by varying the efficiency of the algorithm by 15\% and the mistag rate by 30\%~\cite{BTVPAS}.
The tagging rate is varied according to the flavour of the selected jet as determined from the MC simulation.
This affects the choice of the jets used in the reconstruction of $m_{\rm top}$, and causes the migration of events from one \Bot-tagging multiplicity to another.

A summary of the systematic uncertainties are given in Table~\ref{tab:syst} for the two algorithms, along with their correlations and combined values.
Other sources of uncertainty including template statistics, initial- and final-state radiation, background template shape and normalisation, 
and $\met$ scale, each yield uncertainty on $m_{\rm top}$ of less than $0.5\GeVcc$. They are included in the measurement but are omitted from Table~\ref{tab:syst}.

\begin{table}[htp]
\caption{Summary of the systematic uncertainties (in \!\GeVcc) in the measurement of $m_{\rm top}$, for the two different algorithms, together with their correlations and combined values.}
\label{tab:syst}
\centering
\begin{center}
\begin{tabular}{lcccc}
\hline\hline
Source                  & KINb & AMWT & Correlation factor & Combination\\
\hline
Overall jet energy scale    & +3.1/--3.7  & 3.0       & 1     & 3.1\\
\Bot-jet energy scale & +2.2/--2.5 & 2.5       & 1     & 2.5\\ 
Lepton energy scale & 0.3  & 0.3       & 1     & 0.3\\
Underlying event             & 1.2        & 1.5       & 1     & 1.3\\
Pileup              & 0.9        & 1.1       & 1     & 1.0\\
Jet-parton matching & 0.7        & 0.7 & 1     & 0.7\\
Factorisation scale & 0.7        & 0.6 & 1     & 0.6\\
Fit calibration         & 0.5        & 0.1       & 0     & 0.2\\
MC generator        & 0.9        & 0.2       & 1     & 0.5\\
Parton density functions & 0.4   & 0.6       & 1     & 0.5\\
\Bot-tagging         & 0.3        & 0.5 & 1     & 0.4\\
\hline\hline
\end{tabular}
\end{center}
\end{table}

The fits described above can be turned into a measurement of the \Bot-jet energy scale 
if the top mass is constrained in the fit by using an independent determination. 
To this end, the top mass has been fixed at the current world average value of 
$173.3\pm 1.1\GeVcc$~\cite{topmass_tev} and the JES left free to vary. 
The JES determined in this manner, from a sample composed primarily of \Bot-jets, 
is within 4.8\% of the nominal CMS JES~\cite{JESPAS}. 
The uncertainty on the nominal CMS JES is 3.5--6\% depending on jet $\pt$ and $\eta$.

\subsection{Combination of mass measurements}
\label{sec:mass_combination}
\begin{figure}[htp]
\centering
\includegraphics[width=0.3\textwidth]{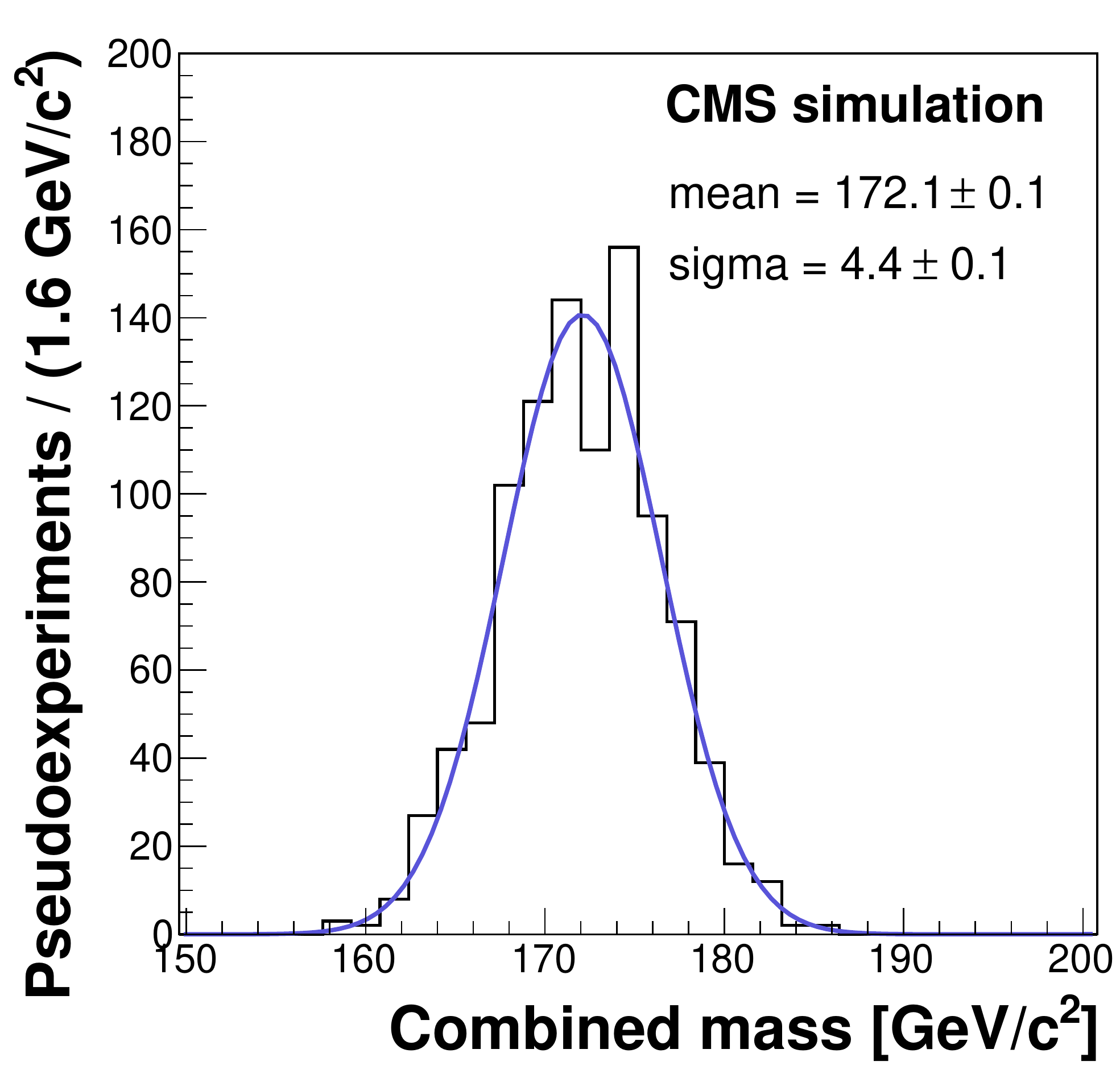}
\includegraphics[width=0.3\textwidth]{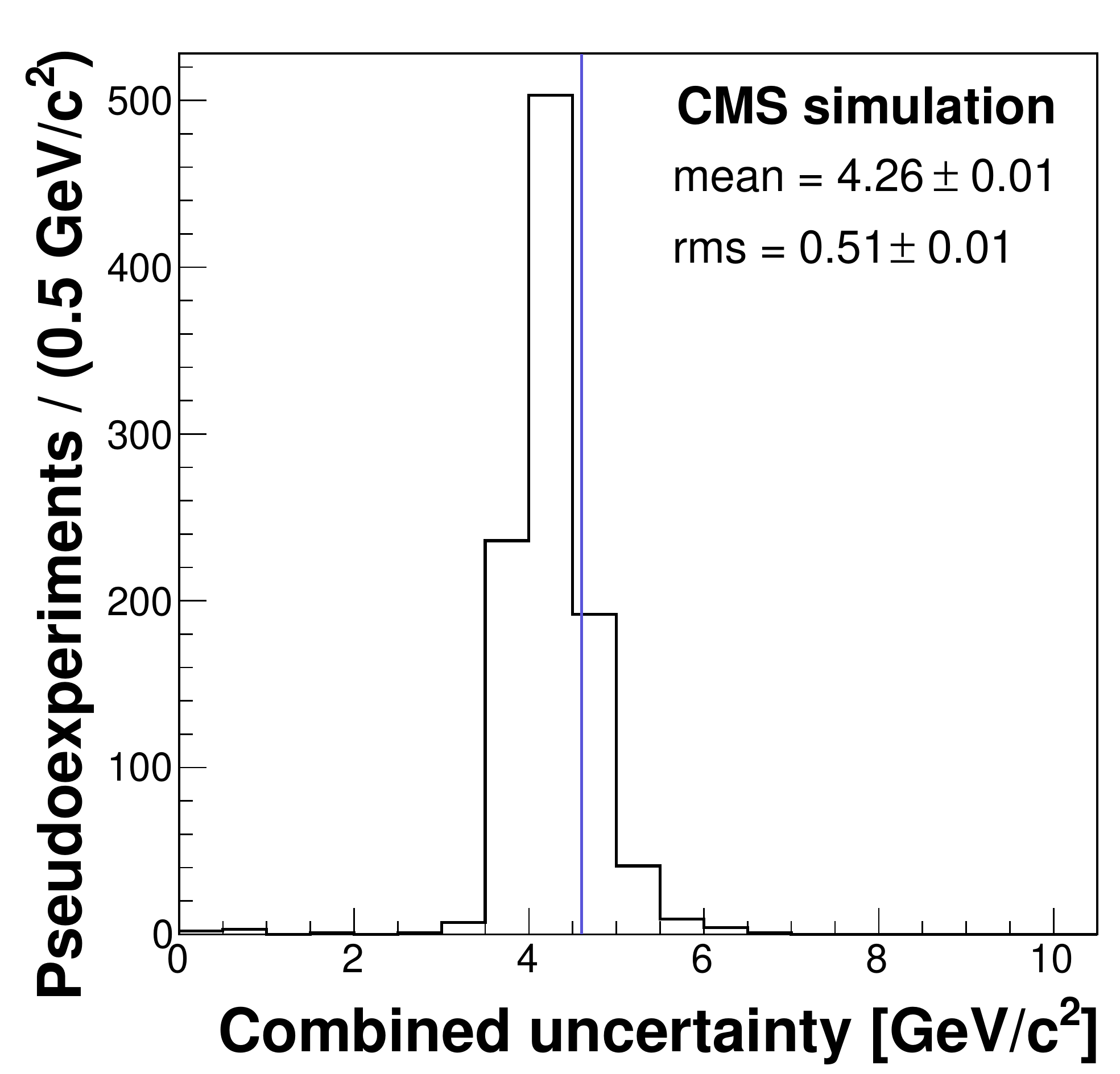}
\caption{Combined 
top quark mass measurements (left) and uncertainties (right)
for pseudo-experiments with $m_{\rm top}=172\GeVcc$.
The result of the fit is shown by the blue line in the left plot.
The statistical uncertainty obtained from the 
combined fit is shown by the vertical blue line
superimposed on the expected uncertainty distribution.}
\label{fig:comb} 
\end{figure}

The BLUE method~\cite{Lyons1988110} is used to combine the KINb and AMWT measurements, with the associated uncertainties and correlation factors.
The statistical correlation between the two methods, which is used to define the contribution of the statistical uncertainties to the error matrix in the combination, 
is determined to be 0.57 from pseudo-experiments with $m_{\rm top}=172\GeVcc$.
In order to check the statistical properties of the combination procedure, the statistical error matrix is computed for each pseudo-experiment and the combination is carried out 
assuming no systematic uncertainties are present. Before proceeding with the combination, the statistical uncertainties are rescaled by the width of the pull distributions so that the pulls with the rescaled uncertainties have an r.m.s. equal to one. The distributions characterising the result of the combination are shown in Fig.~\ref{fig:comb}.
The width of the pull distribution of the combined measurements is very close to unity and 
no further corrections to the statistical uncertainty returned by the combination are needed.

Systematic uncertainties common to the methods are assumed to be 100\% correlated. When individual measurements have asymmetric uncertainties
they are symmetrized before the combination, under the assumption that such asymmetries are not significant and originate from fluctuations in their determination.
The results of the combination are presented in Table~\ref{tab:comb}, along with the individual measurements and the weight they have in the combined result. 

\begin{table}[htp]
\caption{Summary of measured top quark mass for the KINb and AMWT methods with the contributing weights to the combined mass value. 
The $\chi^2/{\rm dof}$ and p-value of the fit are also given.}
\label{tab:comb}
\centering
\begin{center}
\begin{tabular}{ccc} \hline\hline
Method & Measured $m_{\rm top}$ (in \GeVcc) & Weight \\\hline
AMWT      & $175.8\pm4.9\,(\mathrm{stat.})\pm4.5\,(\mathrm{syst.})$ & 0.65 \\
KINb      & $174.8\pm5.5\,(\mathrm{stat.})^{+4.5}_{-5.0}\,(\mathrm{syst.})$ & 0.35 \\\hline
Combined & $175.5\pm4.6\,(\mathrm{stat.})\pm4.6\,(\mathrm{syst.})$  & $\chi^2/{\rm dof}=0.040$ ($\mbox{p-value}=0.84$) \\\hline\hline
\end{tabular}
\end{center}
\end{table}

\section{Summary}
\label{sec:conclusions}
Top quark pair production in proton-proton collisions at \lhcE{7}\ 
has been studied in a data sample corresponding to an integrated luminosity of $36{\rm~pb}^{-1}$ collected 
by the CMS experiment in 2010.
The analysis is based on events with jets, missing transverse energy, and two energetic, well identified, isolated leptons.
Consistent measurements of the \ttbar\ production cross section are obtained from nine final states characterised by combinations of 
lepton flavour ( \eepm, \mmpm, \empm) and number and type of reconstructed jets (one jet, 
two jets with no b-tagged jets, two jets with at least one b-tagged jet).
The combination of these measurements yields 
$\sigma_{{\rm t\bar t}} = 168 \pm 18 ({\rm stat.}) \pm 14 ({\rm syst.}) \pm 7 ({\rm lumi.})~\mrm{pb}$,
in agreement with standard model expectations.
The ratio of production cross sections for \ttbar\ and \dy\
is measured to be $0.175 \pm 0.018 ({\rm stat.})\pm 0.015 ({\rm syst.})$, where the average of the measured 
dielectron and dimuon \dy\ cross sections in  the mass range of 60--$120\GeVcc$ has been used.

The same data sample has been used to perform 
two measurements of the top quark mass using two different kinematic algorithms. The combined result from the two methods is:
$m_{\rm top}=175.5 \pm 4.6 ({\rm stat.}) \pm 4.6({\rm syst.})\GeVcc$. 
This is the first measurement of the top quark mass at the LHC.
With the first year of data-taking, the precision of our top quark mass measurement is already close to that of the Tevatron in the same final state.

\section*{Acknowledgements}
\hyphenation{Bundes-ministerium Forschungs-gemeinschaft Forschungs-zentren}
We wish to congratulate our colleagues in the CERN accelerator departments for the excellent performance of the LHC machine.
We thank the technical and administrative staff at CERN and other CMS institutes.
This work was supported by the Austrian Federal Ministry of Science and Research; the Belgium Fonds de la
Recherche Scientifique, and Fonds voor Wetenschappelijk Onderzoek; the Brazilian Funding Agencies
(CNPq, CAPES, FAPERJ, and FAPESP); the Bulgarian Ministry of Education and Science; CERN;
the Chinese Academy of Sciences, Ministry of Science and Technology, and National Natural Science Foundation of China;
the Colombian Funding Agency (COLCIENCIAS); the Croatian Ministry of Science, Education and Sport;
the Research Promotion Foundation, Cyprus; the Estonian Academy of Sciences and NICPB;
the Academy of Finland, Finnish Ministry of Education and Culture, and Helsinki Institute of Physics;
the Institut National de Physique Nucl\'eaire et de Physique des Particules~/~CNRS,
and Commissariat \`a l'\'Energie Atomique et aux \'Energies Alternatives~/~CEA, France;
the Bundesministerium f\"ur Bildung und Forschung, Deutsche Forschungsgemeinschaft,
and Helmholtz-Gemeinschaft Deutscher Forschungszentren, Germany; the General Secretariat for Research and Technology, Greece;
the National Scientific Research Foundation, and National Office for Research and Technology, Hungary;
the Department of Atomic Energy and the Department of Science and Technology, India;
the Institute for Studies in Theoretical Physics and Mathematics, Iran; the Science Foundation, Ireland;
the Istituto Nazionale di Fisica Nucleare, Italy;
the Korean Ministry of Education, Science and Technology and the World Class University program of NRF, Korea;
the Lithuanian Academy of Sciences; the Mexican Funding Agencies (CINVESTAV, CONACYT, SEP, and UASLP-FAI);
the Ministry of Science and Innovation, New Zealand; the Pakistan Atomic Energy Commission;
the State Commission for Scientific Research, Poland; the Funda\c{c}\~ao para a Ci\^encia e a Tecnologia, Portugal;
JINR (Armenia, Belarus, Georgia, Ukraine, Uzbekistan);
the Ministry of Science and Technologies of the Russian Federation, and Russian Ministry of Atomic Energy;
the Ministry of Science and Technological Development of Serbia;
the Ministerio de Ciencia e Innovaci\'on, and Programa Consolider-Ingenio 2010, Spain;
the Swiss Funding Agencies (ETH Board, ETH Zurich, PSI, SNF, UniZH, Canton Zurich, and SER); the National Science Council, Taipei;
the Scientific and Technical Research Council of Turkey, and Turkish Atomic Energy Authority;
the Science and Technology Facilities Council, UK; the US Department of Energy, and the US National Science Foundation.

Individuals have received support from the Marie-Curie programme and the European Research Council (European Union);
the Leventis Foundation; the A. P. Sloan Foundation; the Alexander von Humboldt Foundation;
the Associazione per lo Sviluppo Scientifico e Tecnologico del Piemonte (Italy); the Belgian Federal Science Policy Office;
the Fonds pour la Formation \`a la Recherche dans l'Industrie et dans l'Agriculture (FRIA-Belgium);
the Agentschap voor Innovatie door Wetenschap en Technologie (IWT-Belgium); and the Council of Science and Industrial Research, India.

\bibliography{auto_generated}

\cleardoublepage\appendix\section{The CMS Collaboration \label{app:collab}}\begin{sloppypar}\hyphenpenalty=5000\widowpenalty=500\clubpenalty=5000\textbf{Yerevan Physics Institute,  Yerevan,  Armenia}\\*[0pt]
S.~Chatrchyan, V.~Khachatryan, A.M.~Sirunyan, A.~Tumasyan
\vskip\cmsinstskip
\textbf{Institut f\"{u}r Hochenergiephysik der OeAW,  Wien,  Austria}\\*[0pt]
W.~Adam, T.~Bergauer, M.~Dragicevic, J.~Er\"{o}, C.~Fabjan, M.~Friedl, R.~Fr\"{u}hwirth, V.M.~Ghete, J.~Hammer\cmsAuthorMark{1}, S.~H\"{a}nsel, M.~Hoch, N.~H\"{o}rmann, J.~Hrubec, M.~Jeitler, W.~Kiesenhofer, M.~Krammer, D.~Liko, I.~Mikulec, M.~Pernicka, H.~Rohringer, R.~Sch\"{o}fbeck, J.~Strauss, A.~Taurok, F.~Teischinger, P.~Wagner, W.~Waltenberger, G.~Walzel, E.~Widl, C.-E.~Wulz
% ...
\vskip\cmsinstskip
% ...
1:~~Also at CERN, European Organization for Nuclear Research, Geneva, Switzerland\\
% ...

\end{sloppypar}
\end{document}